\documentclass{article}
  \usepackage[a4paper,top=3cm,bottom=3cm,left=3cm,right=3cm,marginparwidth=2cm]{geometry}

\usepackage{lmodern}
\usepackage[utf8x]{inputenc} 
\usepackage[T1]{fontenc}
\usepackage[english]{babel}

\usepackage[title]{appendix}
\usepackage{amsmath}
\usepackage{amsthm}
\usepackage{amssymb}
\usepackage{amsfonts}
\usepackage{lipsum}
\usepackage{dsfont}
\usepackage[dvipsnames]{xcolor}
\usepackage{enumitem}
\usepackage{mathabx}
\usepackage{fancyhdr}
\usepackage{stmaryrd}
\usepackage{setspace}
\usepackage{graphicx}
\usepackage{caption}
\usepackage{bm}
\usepackage{subcaption}
\usepackage{capt-of}
\usepackage{mathtools, nccmath}
\usepackage{tikz}
\usepackage{circuitikz}
\usepackage{url}
\usepackage{algpseudocode}
\usepackage{algorithm}
\usepackage[nottoc,notlot,notlof]{tocbibind}
\usepackage{graphicx} 
\usepackage{hyperref} 
\hypersetup{colorlinks=true,breaklinks=true, urlcolor= blue, linkcolor= red, citecolor=blue}
\usepackage{todonotes}
\usepackage{glossaries}

\makeglossaries

\newglossaryentry{hp}
{
    name=HP,
    description={Multivariate nonlinear Hawkes process}
}

\definecolor{mygreen}{HTML}{49be25}

\definecolor{truezero}{HTML}{dbf1d6}
\definecolor{truenonzero}{HTML}{298a44}
\definecolor{falsezero}{HTML}{e0cbe3}
\definecolor{undetectedzero}{HTML}{b796c5}

\theoremstyle{plain}
\newtheorem{theorem}{Theorem}[section]

\newtheorem{proposition}{Proposition}[section]
\newtheorem*{assumption*}{\assumptionnumber}
\providecommand{\assumptionnumber}{}


\theoremstyle{definition}

\numberwithin{equation}{section}

\newcommand\N{\mathbb{N}}

\newcommand\R{\mathbb{R}}

\def\P{\mathbb{P}}
\def\E{\mathbb{E}}

\def\d{\,\mathrm{d}}

\begin{document}
  {
  \title{\textbf{Hawkes Processes with Variable Length Memory: Existence, Inference and Application to Neuronal Activity}}
  \author{Sacha Quayle\thanks{Sorbonne Universit\'e, CNRS, Laboratoire de Probabilit\'es, Statistique et Mod\'elisation, Paris, France}, Anna Bonnet\footnotemark[1], Maxime Sangnier\footnotemark[1]}
  \maketitle
  }
  \maketitle
  \begin{abstract}
  Multivariate Hawkes processes are past-dependant point processes originally introduced to model excitation effects, later extended to a nonlinear framework to account for the opposite effect, known as inhibition.
Motivated by applications in neuroscience, where the memory of a neuron may reset upon firing, we introduce a new class of nonlinear Hawkes processes with variable length memory.
Our model generalises classical Hawkes processes, with or without inhibition, describing the situation where the probability of an event occurring within a given subprocess may depend differently on the history before and after its last event. 
In particular, if the subprocess does not depend on the history before its last event, it is said to have a variable length memory.
Our main contributions are to prove existence of such processes, and to derive a workable likelihood maximisation method, capable of identifying both classical and variable memory dynamics.  
We demonstrate the effectiveness of our approach both on synthetic data, and on a neuronal activity dataset.

  \end{abstract}


\section{Introduction}

In neuroscience, the activity of each neuron is characterised by its membrane potential.
When this potential reaches a certain threshold, the neuron fires and emits a spike, which is extremely brief in time.
It is therefore natural to model a spike as a random event and to represent the entire neuronal activity as a point process consisting of the sequence of spike times \cite{galves_probabilistic_2024}, often using a Hawkes process framework \cite{lambert_reconstructing_2018}. 
In Classical Hawkes process models, the spiking intensity, which caracterises the probability of emitting a spike at a given time, depends on a function of the history, the length of which is deterministically defined to be either \(\infty\) or the size of the function support.
However, after emitting a spike, a neuron's membrane potential often resets to an equilibrium level, causing the spiking intensity to depend on a variable length memory, specifically, on the activity since its last spike.
Classical Hawkes processes do not capture this feature, which motivates the class of models studied in this paper.

Originally, Hawkes processes were univariate self-exciting processes \cite{hawkes_spectra_1971}, but they have since been generalised to a nonlinear multivariate framework \cite{bremaud_stability_1996} to take into account inhibition, where the appearance of events can decrease the probability of occurrence of subsequent events. 
This flexibility allows Hawkes processes to model a large variety of
phenomena, from the time occurrences of earthquakes \cite{ogata_statistical_1988} to social media \cite{rizoiu_tutorial_2017}, criminology \cite{olinde_selflimiting_2020}, ecology \cite{nicvert_using_2024}, or physiological data \cite{loison_unhap_2025}, to cite a few.
Hawkes processes are particularly relevant in applications such as neuroscience \cite{lambert_reconstructing_2018}, where both excitatory and inhibitory effects co-exist.
However, traditional Hawkes process models assume a deterministic memory length: the probability of an event occurring at a given time depends either on the full past history or, when kernel functions have bounded support, on a fixed finite window of past events \cite{sulem_bayesian_2024}.
To better capture the memory reset behaviour observed in neurons, a variant of Hawkes processes with variable length memory was proposed. 
The variable memory structure, in which a neuron's spiking probability depends only on the activity since its last spike, was first introduced in a discrete-time setting in \cite{galves_infinite_2013}. 
However, a formal generalisation of this idea to continuous time, closer in nature to Hawkes processes, was proposed in \cite{galves_modeling_2016,hodara_hawkes_2017}, in a nonlinear and interacting particle system framework on an infinite graph.
However, it is possible some neurons exhibit memory reset, while others do not, which motivates the development of a model that accommodates both fixed and variable memory dynamics, taking into account a possible diversity in memory behaviors.
To the best of our knowledge, only a few works address statistical inference for these models.
In particular, \cite{hodara_hawkes_2017} focus on graph estimation, i.e., inferring the set of presynaptic neurons for each neuron based on its spiking history, while accounting for nonlinearity and memory reset effects.
Other aspects such as parameter estimation, or distinguishing neurons with variable versus fixed memory, remain largely unexplored.

Classical multivariate nonlinear Hawkes process have been widely studied since their introduction in \cite{bremaud_stability_1996} which also prove their existence and stability.
Limit theorems have been established by using renewal theory when the kernel functions are of bounded support \cite{costa_renewal_2020}, and there are mean-field results on the behavior of two neuronal populations \cite{duval_interacting_2022}. 
Regarding statistical inference, in the frequentist setting estimation methods consist of maximum likelihood estimation for exponential kernels \cite{bonnet_inference_2023} and feedforward neural networks \cite{joseph_neural_2024,joseph_nonparametric_2024}, a least-squares approach \cite{bacry_sparse_2020},
nonparametric approaches based on Bernstein-type polynomials \cite{lemonnier_nonparametric_2014} and reproducing kernel Hilbert spaces \cite{bonnet_nonparametric_2025}.
In the neural network community, inference of nonlinear Hawkes processes has also been addressed thanks to specific architectures,
such as long short-term memory recurrent neural networks \cite{mei_neural_2017},
self-attention \cite{zhang_selfattentive_2020}
and transformers \cite{zuo_transformer_2020,meng_interpretable_2024}.

In the Bayesian framework, \cite{sulem_bayesian_2024} proposed a nonparametric estimation procedure for kernel functions with bounded support.
The authors then developed a variational inference procedure \cite{sulem_scalable_2023} in order to reduce the computational cost of their method.
Neural networks have also been used based on Gaussian process priors \cite{pan_selfadaptable_2021}.
Finally, \cite{deutsch_estimating_2025} investigated a parametric inference method based on a new reparametrisation of the process.
Despite these advancements, asymptotic results for the estimators in the inhibitory case are still lacking, highlighting an area for future research.

In this paper, we introduce a generalised model that includes both the classic Hawkes process and the variable length memory Hawkes process, in a finite multidimensional setting.
We study existence conditions to ensure the non-explosion of these processes, generalising spectral radius conditions outlined in \cite{bremaud_stability_1996}.
For the variable length memory model, in which all neurons exhibit memory reset, we establish existence under weaker conditions, when the kernel functions are simply bounded.
We then present a maximum likelihood estimation method for the generalised model using exponential kernel functions, inspired by the methodology presented in \cite{bonnet_inference_2023}, focusing on intervals where the intensity is positive to provide an explicit expression for the log-likelihood, which can be computed recursively.
This enables inference on both the model parameters and the type of memory dynamics involved, implemented in a five-step procedure that includes identifying significant interactions within the entire process, and performing tests to detect classic or variable memory dynamics.
The Markovian properties for the intensities \textcolor{black}{are also useful for computation of the compensator}, which enables us to assess the goodness-of-fit of our model using the time change theorem \textcolor{black}{\cite[Proposition 7.4.IV]{daley_introduction_2008}}.
Our numerical procedure is implemented in Python and is freely available on GitHub \footnote{\href{https://github.com/sachaquayle/hawkesGVM}{https://github.com/sachaquayle/hawkesGVM}}.
We conduct a numerical study on simulated data, consisting of 2 then 10 subprocesses, demonstrating that our method provides accurate estimations and can correctly identify interaction types.
We also apply our method to neuronal activity data to demonstrate the practical relevance of our framework.

Section~\ref{sect:GVM} presents the generalised model, outlines conditions for non-explosion, and derives Markovian properties for the intensities.
Section~\ref{sect:estimation} details the estimation procedure, including maximum likelihood estimation, interaction tests, goodness-of-fit, and the resampling procedure.
Section~\ref{sect:simulations} illustrates the procedure on synthetic datasets, and Section~\ref{sect:application} applies it to neuronal data.
Finally, Section~\ref{sect:conclusion} discusses the limitations of this paper and future perspectives.



\section{Hawkes Processes with Variable Length Memory} \label{sect:GVM}

\subsection{Definitions} \label{sect:def VM hawkes}

In what follows, we denote $\N^* = \N \setminus\{0\}$ the set of integers without $0$, and we consider a multivariate point process $N=(N^1,\dots,N^d)$ of dimension $d \geq 1$, where each component $N^i$, for $1\leq i \leq d$, is a point process on $\R_+$, defined on the Borel $\sigma$-algebra $\mathcal B(\R_+)$.
Each subprocess $N^i$ is characterised by its sequence of event times $(T_k^i)_{k\in \N^*}$, and its law is uniquely determined by its conditional intensity function $\lambda^i$, defined by:
$$\lambda^i(t) = \underset{h\rightarrow 0}{\lim} \frac{\E\left[N^i[t,t+h) | \mathcal F_t\right]}{h}\,,$$
where $(\mathcal F_t)_{t\geq 0}$ is the canonical filtration of $N$. For any $t\geq 0$, we also denote $N^i(t) = \sum_{k\in\N^*} 1_{\{T_k^i \leq t\}}.$
The process $N$ can be seen as a point process on $\R_+$, obtained by superposition of $N^1,\dots,N^d$, with associated event times $(T_{(l)})_{l\in \N^*}$, and $(d_l)_{l\in \N^*}$ the sequence of indexes indicating the dimension to which each event time belongs. \textcolor{black}{Equivalently, $N$ can be viewed as a marked Hawkes process on $\R_+$, where the mark associated with each event specifies the dimension that generated it.}

We consider multivariate Hawkes processes that allow for inhibition, meaning that the occurrence of events can decrease the probability of appearance of future events.
The conditional intensities are thus defined as:
$$\lambda^i(t) = (\lambda^{i*}(t))^+\,,$$
where $x^+ = \max(0,x)$ and $\lambda^{i*}(t)$ is the underlying (possibly negative) intensity of subprocess $N^i$.

A generalised Hawkes process with variable length memory (abbreviated to GVM) is a specific class of multivariate point processes $N=(N^1,\dots,N^d)$ on $\R_+$, where each component $N^i$ has an underlying intensity $\lambda^{i*}$ of the form:
\begin{align} \label{eq:GVM}
    \lambda^{i*}(t)& =\mu_i + \sum_{j=1}^d\left[\int_{[T_{N^i(t)}^i,t[} h_{ij}(t-s)\d N^j(s) + \int_{[0,T_{N^i(t)}^i[} \widetilde{h}_{ij}(t-s)\d N^j(s)\right]  \nonumber \\
    & = \mu_i + \sum_{j=1}^d\left[ \sum\limits_{k\in \N^*,\, T_{N^i(t)}^i\leq T_k^j <t} h_{ij}(t-T_k^j) + \sum\limits_{k\in \N^*,\, T_k^j < T_{N^i(t)}^i} \widetilde h_{ij}(t-T_k^j)\right]\, .
\end{align}
where $\mu_i >0$ is the baseline intensity, which quantifies the rate of event times in the absence of cross-interactions, and $h_{ij},\widetilde h_{ij}: \R_+\rightarrow \R$ are measurable interaction functions describing the influence of $N^j$ on $N^i$.

In the case where $h_{ij}=\widetilde h_{ij}$ for any $1\leq i,j\leq d$, Model~(GVM) resorts to the classical nonlinear Hawkes process model, referred to from now on as Model~(HP), in which the entire past contributes to the intensity.

When $\widetilde h_{ij}=0$ for any $1\leq i,j\leq d$, Model~(GVM) resorts to a Hawkes process model with variable length memory, similar in spirit to the models introduced in \cite{galves_infinite_2013, galves_modeling_2016, hodara_hawkes_2017}, our formulation being in continuous time, involving a finite number of neurons, and defining the conditional intensity through a slightly different form.
In what follows, we refer to this model as Model~(VM).
In this case, the underlying intensity $\lambda^{i*}$ of $N^i$ is:
\begin{align} \label{eq:VM}
    \lambda^{i*}(t)= \mu_i + \sum_{j=1}^d\sum_{k\in\N^*,\, T^i_{N^i(t)} \leq T_k^j<t} h_{ij}(t-T_k^j)\,.
\end{align}
In contrast to Model~(HP), Model~(VM) restricts the memory of subprocess $N^i$ to events occurring after $T^i_{N^i(t)}$, its most recent event, which allows us to model memory reset dynamics, where the memory of each subprocess resets once an event from that subprocess occurs.

Model~(GVM) allows for statistical inference without committing to a specific choice between Models (HP) and (VM), and also allows to account for other memory dynamics than the two previous models, taking into account the possibility of partial memory reset, where some neurons may exhibit memory reset while others do not.


\subsection{Existence results} \label{sect:existence multivariate VM hawkes}

In this section, we investigate conditions ensuring the existence of the process, i.e. preventing explosion.
Given a multivariate point process $N=(N^1,\dots,N^d)$, we say that $N$ is non-explosive, or finite, if, a.s.: $\forall t>0,\, N[0,t] < \infty.$
Equivalently, the process is non-explosive if \cite[p.40]{asmussen_applied_2003}: $\P\left(\underset{l\in\N^*}{\sup} T_{(l)} <\infty \right) = 0$. In the context of multivariate Hawkes processes under Model~(HP), \cite[Theorem 7]{bremaud_stability_1996} showed that non-explosion is guaranteed under the condition:
$$\rho\left( \int_0^\infty |h_{ij}|(t)\d t \right)_{1\leq i,j\leq d} <1\,,$$
where $\rho(A)$ denotes the spectral radius of the matrix $A$.
\textcolor{black}{The authors further established uniqueness, stationarity and stability of the process, which are detailed in their work.} 
A similar existence result can be obtained for Model~(GVM):

\begin{proposition}[\textcolor{black}{Existence for Model (GVM)}] \label{prop:existence GVM}
	Let $h_{ij}, \widetilde h_{ij}: \R_+\rightarrow \R$ be measurable functions.
Assume that:
	$$  \textcolor{black}{\rho  \left( \int_0^{\infty} \max(h_{ij},\widetilde h_{ij})^+(t) \d t \right)_{1\leq i,j\leq d} < 1\,.}
$$
	Then there exists a non-explosive multivariate point process $N$ under Model~(GVM).
\end{proposition}

The result follows by dominating the kernels $h_{ij}$ and $\widetilde h_{ij}$ pointwise by \textcolor{black}{$\max(h_{ij},\widetilde h_{ij})^+$}, and applying a thinning argument based on \cite[Proposition 1]{ogata_lewis_1981}.
The detailed proof is provided in \ref{app:existence GVM}.

By taking $\widetilde h_{ij}= 0$, Proposition~\ref{prop:existence GVM} also gives a spectral radius condition for existence under Model~(VM).
However, a key feature of Model~(VM) is that the memory of each subprocess resets at every event time of that subprocess, which acts as a natural control against explosion.
As a consequence, non-explosion can be ensured under the sole assumption that the interaction functions are bounded.

\begin{theorem}[\textcolor{black}{Existence for Model (VM)}] \label{thm:existence VM}
	Let $h_{ij}: \R_+\rightarrow \R$ be measurable functions.
Assume that the functions $h_{ij}$ are bounded.
	Then there exists a non-explosive multivariate point process $N$ under Model~(VM).
\end{theorem}

The proof relies on dominating the conditional intensity process $\lambda=(\lambda^1,\dots,\lambda^d)$ by a time-continuous Markov chain whose non-explosive behaviour  follows from standard Markov chain theory, and again by applying a thinning argument based on \cite[Proposition 1]{ogata_lewis_1981}.
The detailed proof is provided in \ref{app:existence VM}.

Let us note that this result exclusively concerns Model~(VM), in which the functions $\widetilde h_{ij}$ are null.
In this setting, the assumption required for non-explosion is simply the boundedness of the functions $h_{ij}$.
This condition is mild and is satisfied by most commonly used interaction kernels in practice, such as exponential or compactly supported kernels.
For Model~(GVM), we expect that a similar but weaker result could hold, potentially with conditions primarily involving the functions $\widetilde h_{ij}$, but a formal justification is still under development.


\subsection{Markovian property} \label{sect:recursion VM hawkes}

The goal of this section is to establish Markovian properties for the conditional intensities under Model~(GVM), which will be used later for simulation algorithms and likelihood computation.
In the following, we suppose that the interaction functions are exponential:
\begin{center}
    $h_{ij}(t)= \alpha_{ij}e^{-\beta_{i} t}$ and $\widetilde{h}_{ij}(t) = \widetilde \alpha _{ij} e^{-\widetilde \beta_{i} t}$,
\end{center}
with $\alpha_{ij},\widetilde \alpha_{ij} \in \R$ and $\beta_{i},\widetilde\beta_{i} >0$.

Several variants of the multivariate exponential Hawkes model exist. One common approach is to assume that all interaction functions share a common decay rate $\beta>0$ \cite{bacry_sparse_2020}. 
In this work, we adopt a less restrictive assumption, where the decay rates in each interaction function $h_{ij}$ and $\widetilde h_{ij}$ depend only on the receiving process $N^i$. 
This assumption, also used in the self-exciting case of a 2-dimensional Hawkes process by \cite{ogata_lewis_1981}, allows a balance between model flexibility and computational efficiency, particularly for evaluating the conditional intensities. \textcolor{black}{Biologically, this assumption is also reasonable, as it implies that the rate at which a given process “forgets” an event depends only on the receiving process and not on the source of the event, or whether the triggering event occurred in the recent or distant past.}

In this case, under Model~(GVM), the underlying intensities are, for any $1\leq i \leq d$: 
\begin{equation} \label{eq:underlying intensities multivariate VM}
\lambda^{i*}(t)=\mu_i + \sum_{j=1}^d\left[ \sum\limits_{k\in \N^*,\, T_{N^i(t)}^i\leq T_k^j <t} \alpha_{ij}e^{-\beta_i(t-T_k^j)} + \sum\limits_{k\in \N^*,\, T_k^j < T_{N^i(t)}^i} \widetilde\alpha_{ij}e^{ \widetilde\beta_i(t-T_k^j)}\right].
\end{equation}

We introduce, for any $i\in \{1,...,d\}$:
\begin{center}
\begin{equation} \label{eq:eta,eta_tilde}
\left\{
    \begin{array}{ll}
         \eta^i(t) = \sum\limits_{j=1}^d \sum\limits_{k\in \N^*,\,T_{N^i(t)}^i\leq T_k^j <t} \alpha_{ij} e^{-\beta_i(t-T_k^j)} \\
        \widetilde\eta ^i(t) = \sum\limits_{j=1}^d \sum\limits_{k\in \N^*,\,T_k^j < T_{N^i(t)}^i} \widetilde \alpha _{ij} e^{-\widetilde \beta_i(t-T_k^j)} \\
        \widetilde \eta_{\text{aux}}^i (t) = \sum\limits_{j=1}^d \sum\limits_{k\in \N^*,\,T_{N^i(t)}^i\leq T_k^j <t} \widetilde \alpha_{ij} e^{-\widetilde\beta_i(t-T_k^j)}
    \end{array}
\right.
\end{equation}
\end{center}

where $\eta^i,\widetilde \eta^i$ are such that:
$$\lambda^{i*}(t) = \mu_i + \eta^i (t) + \widetilde \eta^i(t)\,,$$
and $\widetilde \eta_{\text{aux}}^i $ is introduced for computation purposes.

\begin{proposition} \label{prop:recursion multivariate VM hawkes}
Let $N=(N^1,...,N^d)$ be a non-linear multivariate Hawkes process under Model~(GVM) with underlying intensities defined by Equation~\eqref{eq:underlying intensities multivariate VM}.
Let $\eta^i, \widetilde \eta^i$ and $\widetilde \eta_{\text{aux}}^i $ be defined by~\eqref{eq:eta,eta_tilde} for any $1\leq i\leq d$.
We suppose that two different processes cannot jump at the same time, i.e.
for any $1\leq i\neq j \leq d$ and $k_1,k_2\in \N^*$:
$$T_{k_1}^i \neq T_{k_2}^j \text{ a.s.}$$
Then, for any $1\leq i \leq d$:
\[
    \forall t \in (T_{(l)},T_{(l+1)}),\qquad \lambda^{i*}(t) = \mu_i + \left[\eta^i(T_{(l)})+\alpha_{i,d_l}\right]e^{-\beta_i(t-T_{(l)})} + \widetilde\eta^i(T_{(l)}) e^{-\widetilde \beta_i(t-T_{(l)})}\,,
\]
and
\[
    \lambda^{i*}(T_{(l+1)}) =
    \begin{cases}
        \mu_i +  \left[\widetilde\eta^i(T_{(l)}) + \widetilde\eta_{\text{aux}}^i(T_{(l)})+\widetilde \alpha_{i,d_l}\right]e^{-\widetilde\beta_i(T_{(l+1)}-T_{(l)})}  & \text{if }  i= d_{l+1} \\
        \lambda^{i*}(T_{(l+1)}^-) & \text{if } i\neq d_{l+1} \,.
    \end{cases}
\]
\end{proposition}

\vspace{0.2cm}
Proof of Proposition~\ref{prop:recursion multivariate VM hawkes} can be found in \ref{app:recursion multivariate VM hawkes}.

Let us remark that if $\widetilde\alpha_{ij}=0$ for any $1\leq i,j\leq d$, which corresponds to Model~(VM), then, for any $t\in (T_{(l)},T_{(l+1)})$:
$$ \lambda^{i*}(t) = \mu_i + (\lambda^{i*}(T_{(l)}) + \alpha_{i,d_l} - \mu_i)e^{-\beta_i (t-T_{(l)})}\,,$$
thus, strictly between two jump times, the recursive formulas remain the same as in Model~(HP).
The only difference lies in the values $\lambda^{i*}(T_{(l+1)})$ of the intensities precisely at the event times, which reset to $\mu_i$ if $i=d_{l+1}$, reflecting the memory reset.


\section{Estimation} \label{sect:estimation}

\subsection{Maximum likelihood estimation}

As in the previous section, we suppose the kernel functions are exponential:
\begin{center}
    $h_{ij}(t)= \alpha_{ij}e^{-\beta_{i} t}$ and $\widetilde{h}_{ij}(t) = \widetilde \alpha _{ij} e^{-\widetilde \beta_{i} t}$ .
\end{center}
Our objective is to estimate the parameters $(\mu_i,\alpha_{ij},\beta_i,\widetilde\alpha_{ij},\widetilde\beta_i)_{1\leq i,j\leq d}$ using Maximum Likelihood Estimation.
To do so efficiently, it is crucial to compute the log-likelihood recursively.
For any multivariate point process $N=(N^1,\dots,N^d)$, the log-likelihood of the subprocess $N^i$ up to time $t\geq 0$ is given by \cite[Proposition 7.3.III.]{daley_introduction_2003}:
    $$\ell_{\lambda^i}(t) = \sum_{k\geq 0,\, T_k^i <t} \log \lambda^i(T_k^{i-}) - \Lambda^i(t)\,,$$
where $\Lambda^i(t) = \int_0^t \lambda^i(s)\d s$ is the compensator of subprocess $N^i$.
Additionally, the total log-likelihood of the process $N$ is: $\ell_\lambda(t) = \sum_{i=1}^d \ell_{\lambda^i}(t).$

In the case of Hawkes processes under Model~(GVM), the terms $\log \lambda_\theta^i(T_k^{i-}) $ can be computed recursively using Proposition~\ref{prop:recursion multivariate VM hawkes}.
The main difficulty therefore lies in the computation of the compensator, which requires identifying the intervals over which the underlying intensities remain positive.
As detailed in \cite{bonnet_inference_2023}, this computation relies on the monotony of the underlying intensities between consecutive event times, which can be ensured by further assuming, for any $1\leq i\leq d$, that:
$$\widetilde\beta_i = \beta_i\,.$$
This condition implies that the decay rates governing the influence of both recent and distant past events are identical, which is a plausible assumption, and ensures that, between two consecutive event times of $N$, the underlying intensities $\lambda^{i*}$ are increasing or decreasing exponential functions.

In this case, the parameter vector to be estimated is:
$$\theta=(\mu_i, \alpha_{ij}, \beta_i, \widetilde\alpha_{ij})_{1\leq i,j\leq d} = (  \mu,  \alpha,  \beta,  {\widetilde\alpha}) \in (\R_+^*)^d \times \R^{d^2} \times (\R_+^*)^d \times \R^{d^2} \, .$$

In the following, we denote by $\lambda_\theta^{i*}$ and $\Lambda_\theta^i$ respectively the underlying intensity and the compensator of subprocess $N^i$ associated with parameter $\theta$. We also denote by $\lambda_\theta$, $\Lambda_\theta$ \textcolor{black}{and $\ell_\theta$} respectively the intensity, the compensator, \textcolor{black}{and the log-likelihood} of $N$.

\begin{proposition} \label{prop:compensator multivariate}
	Let $N=(N^1,...,N^d)$ be a non-linear multivariate Hawkes process under Model~(GVM) with underlying intensities defined by Equation~\eqref{eq:underlying intensities multivariate VM}. For any $l\in \N^*$ and for any $i\in \{1,...,d\}$, define:
$$ T_{(l), \theta}^{i*} = \min\left(T_{(l+1)},\, \text{inf}\,\{t\geq T_{(l)},\, \lambda_\theta^{i*}(t)\geq 0 \right).$$
    Then, for any $k\in \N^*$ and for any $i\in \{1,...,d\}$:
	$$T_{(l),\theta}^{i*} = \min \left(T_{(l+1)},\, t_{l,\theta}^{i*} \right), $$
	where:
	$$ t_{l,\theta}^{i*} = \left(T_{(l)} + \beta_i^{-1}\log \left( \frac{\mu_i - \lambda_\theta^{i*}(T_{(l)})-\alpha_{i,d_l} }{\mu_i }\right)  1_{\lambda_\theta^{i*}(T_{(l)}^+) <0} \right),$$
	and the compensator of the subprocess $N^i$ is:
	$$\Lambda_\theta^i(t) = \left\{ \begin{array}{cc}
		\mu_i t & \text{ if } t<T_{(1)} \\
		\mu_iT_{(1)}+\sum\limits_{l=1}^{N(t)} J_{l,\theta}^i(t) & \text{ if } t\geq T_{(1)}
	\end{array}
	\right.,$$
	where, for any $1\leq l \leq N(t)$:
	$$J_{l,\theta}^i(t) = \mu_i \left[ \min(T_{(l+1)},t) - T_{(l)}^{i*}\right] +  \frac{\lambda_\theta^{i*}(T_{(l)})-\mu_i + \alpha_{i,d_l} }{\beta_i}\left[ e^{-\beta_i(T_{(l)}^{i*}-T_{(l)})}-e^{-\beta_i (\min(T_{(l+1)},t) -T_{(l)})} \right].
$$
	
\end{proposition}

This result, obtained by using Proposition~\ref{prop:recursion multivariate VM hawkes}, therefore enables a recursive computation of the restart times and makes it possible to evaluate the compensator with the same computational complexity as in the purely self-exciting case.


\subsection{Post hoc estimation of interactions} \label{sect:tests}

\textcolor{black}{In the following estimation procedure, we suppose that we observe $n$ independent realisations $N_1,\dots,N_n$ of a Hawkes process on $[0,T]$.}
Our goal is to identify significant interactions between subprocesses and to test whether these interactions follow classic or variable memory dynamics.
To this end, we test various hypotheses on the coefficients of the interaction matrices $\alpha$ and $\widetilde\alpha$, while accounting for the multiple testing setting through the Benjamini-Hochberg (\textcolor{black}{BH}) procedure \cite{benjamini_controlling_1995} to control the false discovery rate.
We prefer to control the false discovery rate (FDR) instead of the probability of making at least one false discovery (FWER) for two reasons.
First, our method is intended to be applied to situations were the number of hypotheses to test is relatively large, in which FWER is known to be too conservative.
Second, for the FDR controlling procedure is used for interaction detection before re-estimation, we can allow without damage a few false discoveries (in that case, re-estimation would lead to low intensity interactions).
We are exactly in the motivating situation of FDR, where we want to discover promising avenues of investigation (i.e. estimation) while controlling the proportion of wrong leads.

In each testing setting, the approach consists of three steps: construct confidence intervals for the parameters of interest, compute the corresponding $p$-values for the relevant null hypotheses, then apply the \textcolor{black}{BH} procedure to select significant effects.

\textcolor{black}{Two types of confidence intervals can be considered.}
The first type relies on the asymptotic normality of the estimators, \textcolor{black}{using the Student distribution or the Hotelling distribution}. This assumption is true in the case of purely exciting interactions \textcolor{black}{but is not guaranteed theoretically under Model~(GVM). 
For Model~(GVM), asymptotic normality is only conjectured and can be checked empirically (see Section~S1 of the Supplementary Material). This is discussed in the following section.}
The second type of confidence intervals that can be considered is empirical confidence intervals, constructed directly from the empirical distribution of the estimators across all realisations. \textcolor{black}{More precisely, for a significance level $\eta \in (0, 1)$, we use the $(\lfloor \frac{\eta}{2}n \rfloor)$ and $(\lceil(1 - \frac{\eta}{2})n\rceil)$-th smallest values.
For example, a confidence interval for $\alpha_{ij}$ is 
\begin{center}
    $\left [\alpha_{ij}^{(\lfloor \frac{\eta}{2} n \rfloor)},\alpha_{ij}^{(\lceil \left( 1-\frac{\eta}{2}\right) n \rceil)}\right], $
\end{center}
where $(\alpha_{ij}^{(k)})_{1\leq k\leq n}$ is the ordered sequence of estimations of $\alpha_{ij}$, with the convention $\alpha_{ij}^{(0)}=-\infty$.}

Our experience is that existence of interactions is often well detected even in misspecified models (such as a simple nonlinear exponential model).
Armed with this a priori, we propose to test first existence of interactions, then its nature (with or without memory reset). 
The full testing procedure consists of:
\begin{center}
\fbox{\begin{minipage}{\textwidth}
\textbf{Estimation procedure under Model (GVM):}
\begin{enumerate}
    \item \textbf{Step~1} (Initial estimation): Compute maximum likelihood estimators \(\hat \theta_1, \dots, \hat \theta_n\) under Model~(GVM) for each realisation.
        
    \item \textbf{Step~2} (Estimation of interactions): Estimate interacting subprocesses by testing, for any \(1 \le i, j \le d\), the null hypotheses $(\alpha_{ij}, \widetilde{\alpha}_{ij}) = (0, 0)$ (see Test~1) with levels corrected by the \textcolor{black}{BH} procedure.
    
    \item \textbf{Step~3} (Re-estimation): Set insignificant coefficients of the matrix \(\alpha\) to zero and re-compute \(\hat \theta_1, \dots, \hat \theta_n\).
    
    \item \textbf{Step~4} (Nature of interactions): Estimate the nature of interactions by testing, for any \(1 \le i, j \le d\), the null hypotheses $\widetilde\alpha_{ij} = 0$ (see  Test~2) and $\alpha_{ij} = \widetilde\alpha_{ij}$ (see Test~3)  with levels corrected by the \textcolor{black}{BH} procedure.
    
    \item \textbf{Step~5} (Final estimation): Compute final estimators \(\hat \theta_1, \dots, \hat \theta_n\) for each realisation, \textcolor{black}{or final estimator $\hat\theta$ over all realisations,} considering the test results from Step~4.
\end{enumerate}
\end{minipage}}
\end{center}

\textcolor{black}{In Step~2, Test~1 is performed for the joint hypotheses $(\alpha_{ij}, \widetilde{\alpha}_{ij}) = (0, 0)$ in order to directly detect the absence of interaction between subprocesses. This approach also reduces the number of multiple tests, as it requires $d^2$ tests instead of $2d^2$ if we were to test each parameter separately.}

\textcolor{black}{In Step~5, the final estimation can be obtained using two different methods.
The first approach is by averaging the estimators $(\hat\theta_1,\dots,\hat\theta_n)$ obtained for each realisation.
The second consists in maximising the sum of the log-likelihoods over all realisations, $\sum_{k=1}^n \ell_\theta^k$.
The advantages of both methods are discussed in the following section.}

\begin{center}
\fbox{\begin{minipage}{\textwidth}
\textbf{Test~1:} Null hypothesis $(\alpha_{ij},\widetilde\alpha_{ij}) = (0,0)$.
\begin{enumerate}
    \item For estimations \(\widehat\alpha_{ij}^k\) and \(\widehat{\widetilde\alpha}_{ij}^k\) (of \(\alpha_{ij}\) and \(\widetilde \alpha_{ij}\)) computed on the \(k^\text{th}\) repetition,
    define $\gamma_{ij}^k = \begin{pmatrix} \widehat\alpha_{ij}^k \\ \widehat{\widetilde\alpha}_{ij}^k \end{pmatrix}$.
    \item Set:
	\begin{itemize}
		\item $\overline{\gamma_{ij}}  := \frac1n \sum_{k=1}^n \gamma_{ij}^k$;
		\item $S_{ij}= \frac{1}{n-1} \sum_{k=1}^n \left(  \gamma_{ij}^k -  \overline{\gamma_{ij}} \right)
 \left(  \gamma_{ij}^k -  \overline{\gamma_{i,j}} \right)^\top$.
	\end{itemize}
	\item Defining the Hotelling \(t\)-squared statistic \(t^2_{ij} = n \overline{\gamma_{ij}}^\top\, S_{ij}^{-1} \,\overline{\gamma_{ij}}\),
	compute:
	\begin{itemize}
	    \item the \(p\)-value for an asymptotic confidence interval: \(\eta_0 = 1-F(t^2_{ij})\), where \(F\) is the cumulative distribution function of the Hotelling \(T\)-squared distribution with parameters \(2\) and \(n-1\);
	    \item the \(p\)-values for an empirical confidence interval: $\eta_0 = 2 \frac{\min(k_0^+,k_0^-)}{n}$, where $k_0^+$ (resp. $k_0^-$) is the number of strictly positive (resp. negative) estimations $\alpha_{ij}^k$, $k\in\{1,\dots,n\}$, and $\widetilde\eta_0 = 2 \frac{\min(\widetilde k_0^+,\widetilde k_0^-)}{n}$, where $\widetilde k_0^+$ (resp. $\widetilde k_0^-$) is the number of strictly positive (resp. negative) estimations $\widetilde\alpha_{ij}^k$, $k\in\{1,\dots,n\}$. 
	\end{itemize}
\end{enumerate}
\end{minipage}}
\end{center} 

\begin{center}
\fbox{\begin{minipage}{\textwidth}
\textbf{Test~2:} Null hypothesis $\widetilde\alpha_{ij} = 0$.
\begin{enumerate}
	\item For any $1\leq i,j\leq d$ where an interaction was detected in Test~1, set:
	\begin{itemize}
		\item $\overline{\widetilde \alpha}_{ij}= \frac1n \sum_{k=1}^n\widehat{\widetilde{\alpha}}_{ij}^k$;
		\item $\widetilde S_{ij} = \frac{1}{n-1} \sum_{k=1}^n \left(\widehat{\widetilde{\alpha}}_{ij}^k-\overline{\widetilde \alpha}_{ij}\right)^2$.
	\end{itemize}
	\item Defining the statistic $t_{ij}=\overline{\widetilde \alpha}_{ij}\sqrt{\frac{n}{\widetilde S_{ij}  }}$, compute the $p$-value:
	\begin{itemize}
	    \item for an asymptotic confidence interval: $\eta_0 = 2(1-F(t_{ij}))$, where $F$ is the cumulative distribution function of the Student distribution with parameter $n-1$;
	    \item for an empirical confidence interval: $\eta_0 = 2 \frac{\min(k_0^+,k_0^-)}{n}$, where $k_0^+$ (resp. $k_0^-$) is the number of strictly positive (resp. negative) estimations $\alpha_{ij}^k$, $k\in\{1,\dots,n\}$.
	\end{itemize}
\end{enumerate}
\end{minipage}}
\end{center}

\begin{center}
\fbox{\begin{minipage}{\textwidth}
\textbf{Test~3:} Null hypothesis $\alpha_{ij} = \widetilde\alpha_{ij}$.
\begin{enumerate}
	\item For any $1\leq i,j\leq d$ where an interaction was detected in Test~1, set:
	\begin{itemize}
		\item $\overline{\alpha}_{ij}= \frac1n \sum_{k=1}^n\widehat{\alpha}_{ij}^k$ and $\overline{\widetilde \alpha}_{ij}^{(n)} = \frac1n \sum_{k=1}^n\widehat{\widetilde{\alpha}}_{ij}^k$;
		\item $S_{ij} = \frac{1}{n-1} \sum_{k=1}^n \left(\widehat{\alpha}_{ij}^k - \widehat{\widetilde{\alpha}}_{ij}^k-\overline{\alpha}_{ij}+ \overline{\widetilde \alpha}_{ij}\right)^2$.
	\end{itemize}
	\item Defining the statistic $t_{ij} =\left(\overline{\alpha}_{ij}- \overline{\widetilde \alpha}_{ij}\right) \sqrt{\frac{n}{S_{ij}   }} $, compute the $p$-value:
	\begin{itemize}
	    \item for an asymptotic confidence interval: $\eta_0 = 2(1-F(t_{ij}))$, where $F$ is the cumulative distribution function of the Student distribution with parameter $n-1$;
	    \item for an empirical confidence interval: $\eta_0 = 2 \frac{\min(k_0^+,k_0^-)}{n}$, where $k_0^+$ (resp. $k_0^-$) is the number of strictly positive (resp. negative) estimations $\alpha_{ij}^k-\widetilde\alpha_{ij}^k$, $k\in\{1,\dots,n\}$.
	\end{itemize}
\end{enumerate}
\end{minipage}}
\end{center}


\subsection{Goodness-of-fit} \label{sect:goodness_fit}

When trying to model certain phenomena through a Hawkes process model, it is important to be able to verify the goodness-of-fit of an estimated model obtained.
The computation of the compensator in Proposition~\ref{prop:compensator multivariate} enables the use of the Time Change Theorem for inhomogeneous Poisson processes  \cite[Proposition 7.4.IV]{daley_introduction_2008}, which states that a sequence of times $(T_{(l)})_{l\in \N^*}$ is a realisation of a point process with conditional intensity $\lambda$ if and only if the sequence of transformed times $(\Lambda(T_{(l)}))_{l\in \N^*}$ is a realisation of a unit-rate Poisson process.

Given $n$ realisations $N_1,\dots,N_n$ with event times denoted $\left(T_{(l), k}\right)_{l\in \N^*}$ for any $1\leq k\leq n$, a first approach would therefore be to compute, for each realisation $N_k$, an estimator $\hat \theta_k$, and to test whether $(T_{(l), k})_{l\in \N^*}$ is a realisation of a point process with conditional intensity $\lambda_{\hat\theta_k}$. 
This can be done using \textcolor{black}{an equality} test between the empirical distribution $(\Lambda_{\hat\theta_k,k}(T_{(l+1), k}) - \Lambda_{\hat\theta_k,k}(T_{(l), k}))_{l\in\N^*}$ and the unit exponential distribution, where $\Lambda_{\hat\theta_k,k}$ is the compensator associated to the conditional intensity $\lambda_{\hat\theta_k,k}$ of $N_k$ with parameter $\hat\theta_k$.
As deeply discussed in \cite{reynaud-bouret_goodnessoffit_2014}, it is explained that doing the \textcolor{black}{goodness-of-fit} test with the same sequence of event times as the one used for building the MLE will tend to over-estimate the $p$-value, whilst doing the \textcolor{black}{test} with an independent sample of event times will tend to under-estimate the $p$-value. 
Therefore, to correct the calculated $p$-values, a method based on resampling is proposed by the authors \cite[Test 4]{reynaud-bouret_goodnessoffit_2014}. 
In this procedure, the event times of each realisation are transformed using the compensator, a random subsample of the realisations is drawn then concatenated using the appropriate shifts, and the test is performed between the resulting times and the unit-rate exponential distribution. The full procedure is decribed below.

\begin{center}
\fbox{\begin{minipage}{\textwidth}
	\textbf{Resampling procedure for goodness-of-fit:}
	\begin{enumerate}
		\item Observe $n$ realisations $N_1,\dots,N_n$ of multivariate Hawkes processes on $[0,T]$.
The event times for each realisation are denoted $\left(T_{(l), k}\right)_{1\leq l\leq n_k}$ for any $1\leq k \leq n$, where $n_k := N_k[0,T]$.
		
		\item Estimate the \textcolor{black}{final} parameter $\hat\theta$ using the $n$ realisations.
		
		\item For any $1\leq k \leq n$, apply the time-transformation to obtain the points $\widehat{\mathcal N}_k = \left(\Lambda_{\hat\theta, k}(T_{(l), k})\right)_{1 \leq l \leq n_k}$ on $[0,\Lambda_{\hat\theta, k}(T)]$.
		
		\item Randomly select a subsample $S = \{i_1, \dots, i_{p_n}\}$ from $\{1, \dots, n\}$ with cardinality $p_n$ such that $p_n/n \underset{n\rightarrow \infty}{\longrightarrow} 0$ (for example, $p_n=\lfloor\sqrt{n}\rfloor$ or $p_n= \lfloor n^{2/3})\rfloor $.
		
		\item Concatenate the $p_n$ processes $\widehat{\mathcal N}_k$ for $k\in S$ to form $\widehat{\mathcal N}^{c,p_n}$.
		
		\item Denote $M = p_n^{-1} \sum_{k\in S} \Lambda_{\hat\theta, k}(T) $, and fix $0<\theta < M$ (for example, $\theta = 0.9M$).
		
		\item Use a \textcolor{black}{goodness-of-fit} test to check whether the times $\{T\in \widehat{\mathcal N}^{c,p_n},\, T\leq p_n\theta\}$ are drawn from a homogeneous Poisson process with intensity 1.
	\end{enumerate}
\end{minipage}}
\end{center}


\section{Numerical results on synthetic datasets} \label{sect:simulations}

\subsection{Simulation procedure}

We consider non-linear Hawkes processes under Model~(GVM) with conditional intensities defined by Equation~\eqref{eq:GVM}, with $h_{ij}(t)=\alpha_{ij}e^{-\beta_i t}$ and $\widetilde h_{ij}(t)=\widetilde\alpha_{ij}e^{-\beta_i t}$, such that the parameters $ \alpha :=(\alpha_{ij})_{1\leq i,j\leq d}, \,   \beta := (\beta_i)_{1\leq i\leq d},$ and $ {\widetilde \alpha} := (\widetilde \alpha_{ij})_{1\leq i,j\leq d}$ satisfy the existence conditions presented in Section~\ref{sect:existence multivariate VM hawkes}.
We denote $\theta = (  \mu,  \alpha, \beta, {\widetilde\alpha})$.

The subsequent sections analyse parameter estimation under each model using simulated data, with an emphasis on estimation under Model~(GVM).
For each scenario considered, 25 independent realisations are generated via the thinning method, ensuring each dataset contains 5000 event times.
Parameter estimation under Model~(GVM) follows the 5-Step procedure described in Section~\ref{sect:tests}, with parameters estimated by minimising the negative log-likelihood computed via Proposition~\ref{prop:recursion multivariate VM hawkes} and Proposition~\ref{prop:compensator multivariate}, using the L-BFGS method from the scipy.optimize.minimize library \cite{virtanen_scipy_2020}.
Parameter estimation under Models (HP) and (VM) simply follows Steps~1 to 3, by replacing the null hypothesis in Test~1 by $\alpha_{ij}=0$.

For all of the considered tests, we use asymptotic normality-based confidence intervals \textcolor{black}{since the number of realisations in our dataset is limited.} As explained in Section~\ref{sect:tests}, it is also possible to use empirical confidence intervals, \textcolor{black}{however they require a large number of realisations to be reliable. 
Additional comparisons between empirical and asymptotic confidence intervals, as well as a study of the empirical distributions to support the asymptotic normality conjecture, are provided in Section~S1 of the Supplementary Material.} 

\textcolor{black}{In Step~5, as discussed in Section~\ref{sect:tests}, the final estimations can be computed in two different ways. 
Estimations for each realisation are done in order to compare and produce the boxplots of the relative squared errors.
However, to assess goodness-of-fit, the final estimator $\hat\theta$ is computed by maximising the sum of the log-likelihoods, instead of averaging the estimations for each realisation.
This second approach is preferred to obtain a final estimator, as it ensures that the resulting parameters define a conditional intensity that remains non-negative at the observed event times.
Additional details on the comparison of the different estimators are provided in Section~S2 of the Supplementary Material.}

\textcolor{black}{Finally, the $p$-values are obtained using the resampling procedure described in Section~\ref{sect:goodness_fit}. The equality test considered here is the Cramer-von-Mises (CvM) test, using the CvM test implemented in the scipy.cramervonmises library. This test is based on the integrated squared difference between the empirical and theoretical cumulative distribution functions, and is generally more powerful than the usual Kolmogorov-Smirnov (KS) test. A detailed comparison between the KS and CvM tests is provided in Section~S3 of the Supplementary Material to further support this choice.}


\subsection{Estimation on bivariate processes}

We first consider an exponential bivariate Hawkes process with parameters
\[
    \mu = \begin{pmatrix} 0.7 \\ 1 \end{pmatrix},
    \quad
    \alpha = \begin{pmatrix} 0.2 & 0 \\ -0.6 & 1.2 \end{pmatrix}
    \quad \text{and} \quad
    \beta = \begin{pmatrix} 3 \\ 2  \end{pmatrix}.
\]
From it, two different scenarios are distinguished:
the first one, named Scenario~(HP), corresponds to simulating under Model (HP), that is taking
\[
    \widetilde \alpha = \begin{pmatrix} 0.2 & 0 \\ -0.6 & 1.2 \end{pmatrix}.
\]
The second one, named Scenario~(VM), corresponds to simulating under Model (VM), that is taking \(\widetilde \alpha = 0\).

\begin{figure}[ht!]
\centering
    \includegraphics[width=0.55\textwidth]{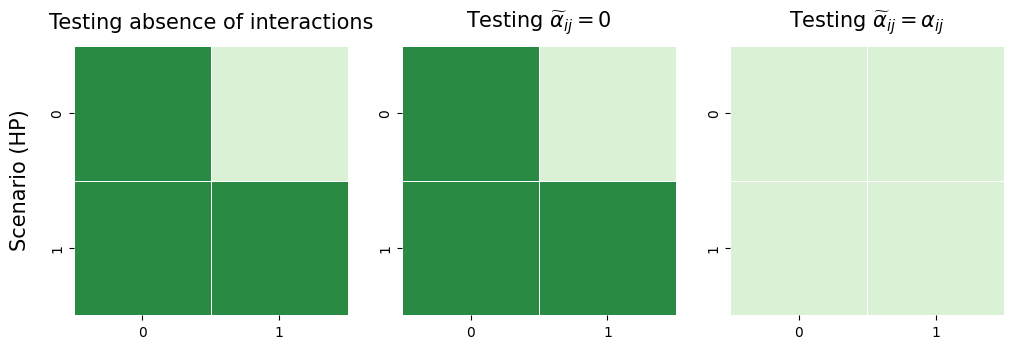}\\[1ex] 
    \includegraphics[width=0.55\textwidth]{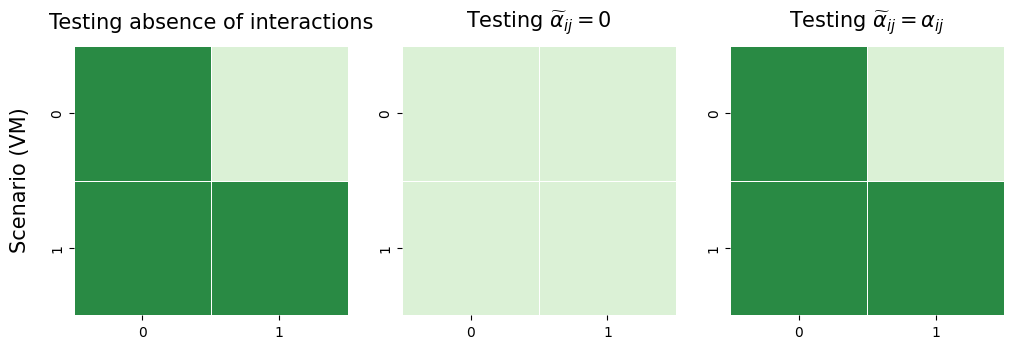}
     \captionsetup{justification=centering}
    \caption{Test results from estimations under (GVM) for Scenarios~(HP) and (VM).
Colours mean true positive \textcolor{truezero}{\(\blacksquare\)} (detected 0 or equal value),
true negative \textcolor{truenonzero}{\(\blacksquare\)} (detected non-null value or non-equal value),
false positive \textcolor{falsezero}{\(\blacksquare\)} (non-null value set to 0 or non-equal value set equal),
false negative \textcolor{undetectedzero}{\(\blacksquare\)} (undetected 0 or equal value).
}
    \label{fig:tests_dim2}
\end{figure}

\begin{figure}[ht!]
\centering
    \includegraphics[width=0.76\textwidth]{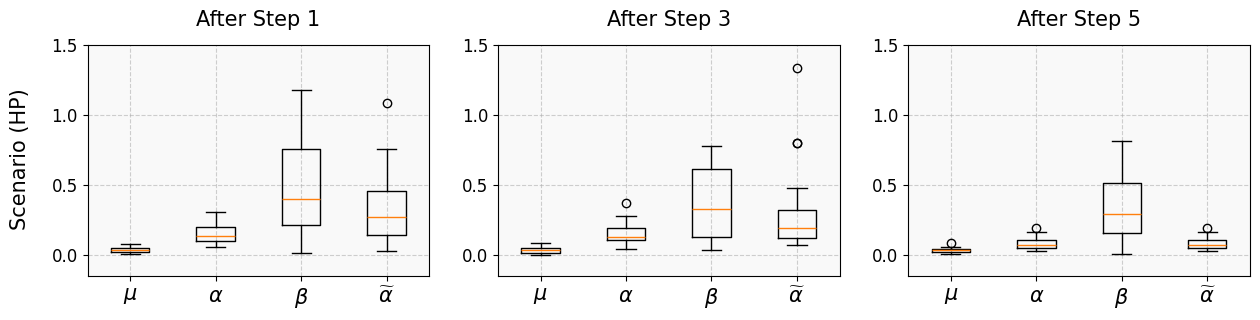}\\[1ex] 
    \includegraphics[width=0.76\textwidth]{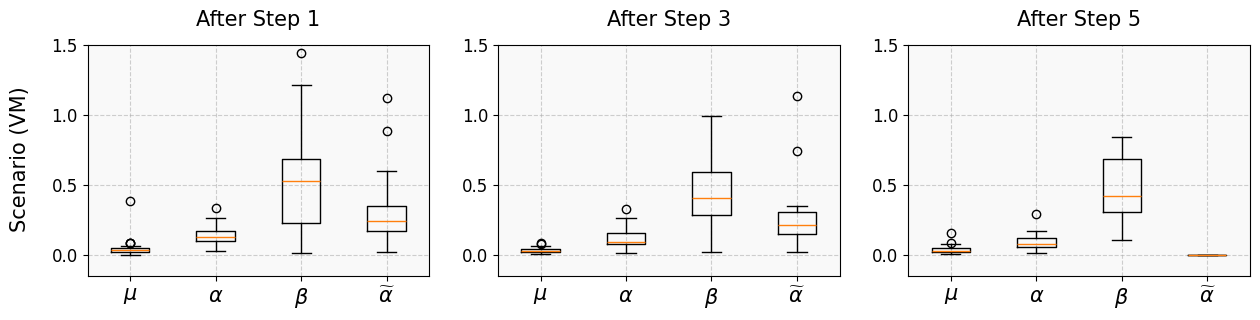}
 \captionsetup{justification=centering}
\caption{Boxplots of the relative squared error for each group of parameters ($ \mu,\,  \alpha,\,  \beta$ and $ {\widetilde\alpha}$) under (GVM) for Scenarios~(HP) and (VM), after Steps~1, 3 and 5 of the estimation procedure.} \label{fig:errors_dim2}
\end{figure}

For each scenario, we display the outcomes from Tests~1 to 3, and the relative squared errors for each group of parameters ($ \mu,\,  \alpha,\,  \beta$ and $ {\widetilde\alpha}$) after Steps~1, 3 and 5, by considering vector norms.
The results show that Model~(GVM) achieves accurate parameter estimation in both scenarios.
As illustrated in Figure~\ref{fig:tests_dim2}, Tests~1 to 3 successfully recover the support and the correct interaction type.
Figure~\ref{fig:errors_dim2} illustrates that the relative errors progressively decrease after each test, highlighting the effectiveness of our five-step procedure. Most of the error reduction occurs between Steps~3 and 5, primarily affecting the estimates of $\beta$ and $\widetilde \alpha$. The error on $\alpha$ remains relatively stable throughout the estimation procedure, which validates our procedure consisting of testing absence of interactions first. 
Although the errors on $\widetilde\alpha$ are higher than those on $\alpha$, it is guessed that, since \(\widetilde \alpha\) represents distant past influence, its contribution is substantially smaller than recent past when the kernel is exponential, and is thus more difficult to estimate.

In Table~\ref{table:pvalues_dim2}, we also display the average $p$-value from the outcomes of the goodness-of-fit test for Model (GVM), by applying the resampling procedure \textcolor{black}{50} times.
These tests are performed with the \textcolor{black}{final estimation obtained after Step~5, computed over all realisations}.
In comparison, \(p\)-values are also provided for the true parameters as well as with estimations under Models (HP) and (VM), where the final estimates are taken after Step~3.
These results indicate that Model~(GVM) provides consistently high $p$-values, similar to those obtained with the true parameters in both scenarios, demonstrating its flexibility and ability to fit the data well.
Model~(HP) performs well when data is from Scenario~(HP) but fails when the data comes from Scenario~(VM), as reflected by lower $p$-values.
Similarly, Model~(VM) fits (VM) data accurately but performs poorly for (HP) data.
This highlights the limitations of each specific model and the robustness of the generalised Model~(GVM).

\begin{table}[ht!]
\centering
\begin{tabular}{|c|ccc|ccc|}
\hline
Scenario & \multicolumn{3}{c|}{(HP)} & \multicolumn{3}{c|}{(VM)} \\
\hline
Model & (HP) & (VM) & (GVM) & (HP) & (VM) & (GVM) \\
\hline
True  & \textcolor{black}{0.49} & \textcolor{black}{6e-8} & \textcolor{black}{0.49}  & \textcolor{black}{1e-8} & \textcolor{black}{0.48} & \textcolor{black}{0.51} \\
\hline
Estimations  & \textcolor{black}{0.47} & \textcolor{black}{0.27} & \textcolor{black}{0.5} & \textcolor{black}{0.13} & \textcolor{black}{0.43} & \textcolor{black}{0.49}  \\
\hline
\end{tabular}
\captionsetup{justification=centering}
\caption{Average $p$-values \textcolor{black}{obtained after resampling realisations} from Scenarios~(HP) and (VM).}  \label{table:pvalues_dim2}
\end{table}


\subsection{Estimation on 10-dimensional processes}

\begin{figure}[ht!]
\centering
\includegraphics[width=0.7\linewidth]{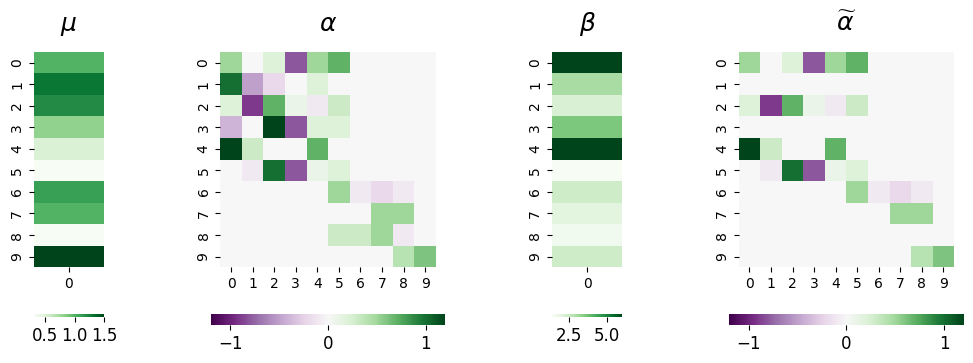}
 \captionsetup{justification=centering}
\caption{\textcolor{black}{Heatmaps of true parameters $ \mu,\,  \alpha, \,  \beta$ and $ {\widetilde\alpha}$.}} \label{fig:params10}
\end{figure}

We consider $10$-dimensional parameters $ \mu,\,  \alpha,  \beta$ and $ {\widetilde\alpha}$ whose heatmaps are displayed in Figure~\ref{fig:params10} and whose exact values can be found on GitHub.
\textcolor{black}{The value for $\alpha$ has been chosen to include sparsity and both excitatory and inhibitory interactions.}
The chosen value for $ {\widetilde\alpha}$ reflects that all subprocesses exhibit classic Hawkes behaviour, except for Subprocesses~1, 3 and 8 which have a variable length memory.
We consider four scenarios listed below: corresponding to scenarios under Models~(HP), (VM) and (GVM), and an additional scenario where the values of $\beta$ are multiplied by a factor 10 (thus indicating the rate at which the process forgets past events).
\begin{itemize}
    \item Scenario~(HP): Parameters $ \mu,\,  \alpha,  \beta$ are given by Figure~\ref{fig:params10} and $ {\widetilde\alpha}= \alpha$.
    \item Scenario~(VM): Parameters $ \mu,\,  \alpha,  \beta$ are given by Figure~\ref{fig:params10} and $ {\widetilde\alpha}=0$.
    \item Scenario~(GVM): Parameters $ \mu,\,  \alpha,  \beta$ and $ {\widetilde\alpha}$ are given by Figure~\ref{fig:params10}.
    \item Scenario~(HP-short-mem): Parameters $ \mu,\,  \alpha,  \beta$ are given by Figure~\ref{fig:params10}, $ \beta$ is multiplied by a factor 10, and $ {\widetilde\alpha}= \alpha$.
\end{itemize}

\begin{figure}[ht!]
\centering
    \includegraphics[width=0.6\textwidth]{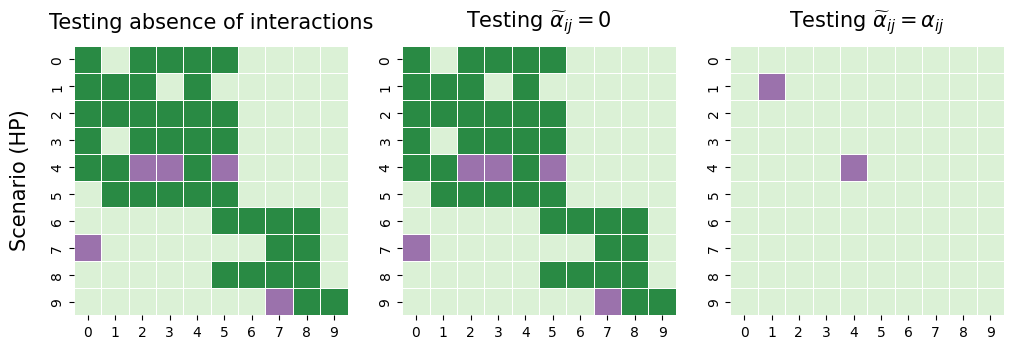}\\[1ex] 
     \includegraphics[width=0.6\textwidth]{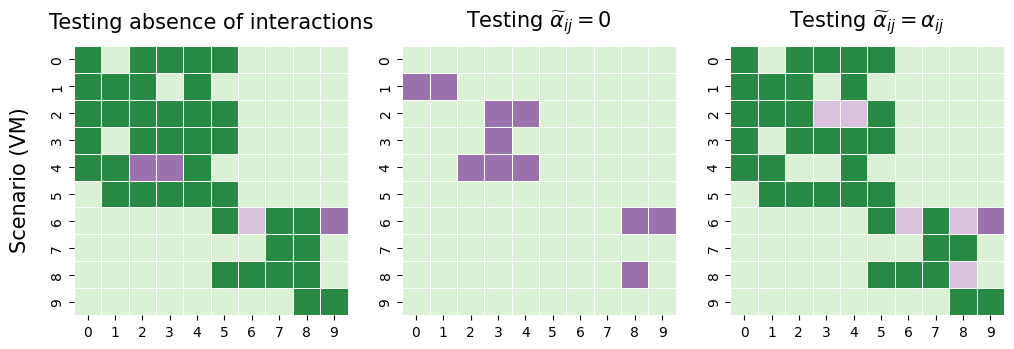}\\[1ex] 
      \includegraphics[width=0.6\textwidth]{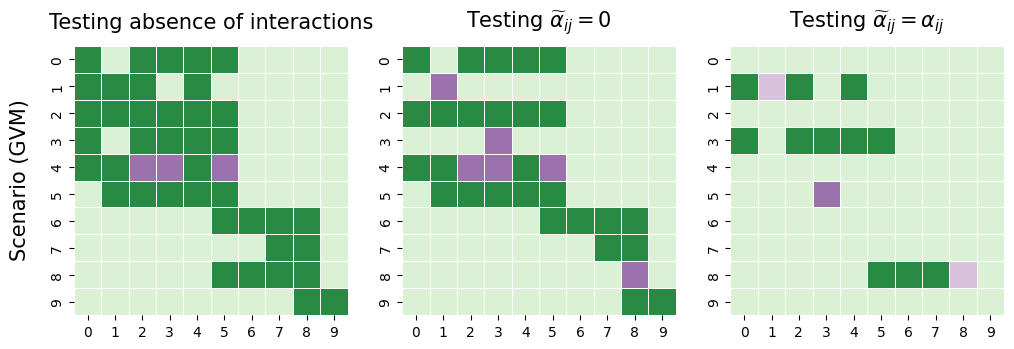}\\[1ex] 
       \includegraphics[width=0.6\textwidth]{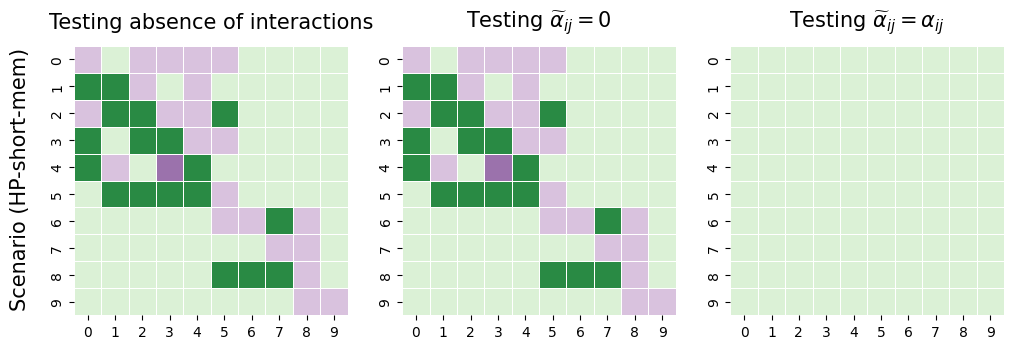}\\[1ex] 
 \captionsetup{justification=centering}
\caption{Test results from estimations under (GVM) for Scenarios~(HP), (VM), (GVM) and (HP-short-mem).
Colours mean true positive \textcolor{truezero}{\(\blacksquare\)} (detected 0 or equal value),
true negative \textcolor{truenonzero}{\(\blacksquare\)} (detected non-null value or non-equal value),
false positive \textcolor{falsezero}{\(\blacksquare\)} (non-null value set to 0 or non-equal value set equal),
false negative \textcolor{undetectedzero}{\(\blacksquare\)} (undetected 0 or equal value).
} \label{fig:tests_dim10}
\end{figure}

\begin{figure}[ht!]
    \centering

    \begin{subfigure}[t]{0.27\textwidth}
        \centering
        \includegraphics[width=\linewidth]{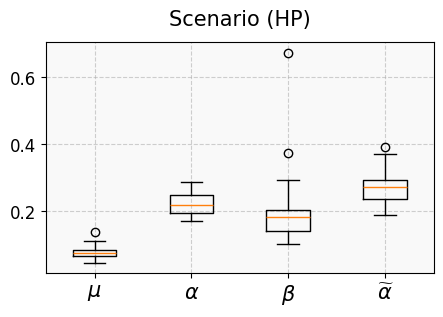}
    \end{subfigure}
    \hspace{0.02\textwidth}
    \begin{subfigure}[t]{0.27\textwidth}
        \centering
        \includegraphics[width=\linewidth]{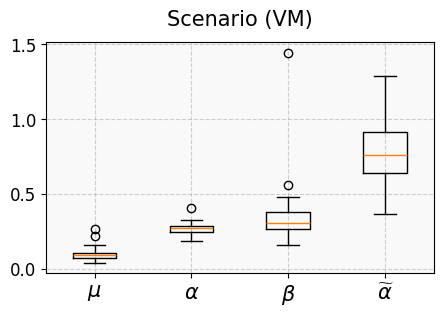}
    \end{subfigure}

    \vspace{0.15cm} 

    \begin{subfigure}[t]{0.27\textwidth}
        \centering
        \includegraphics[width=\linewidth]{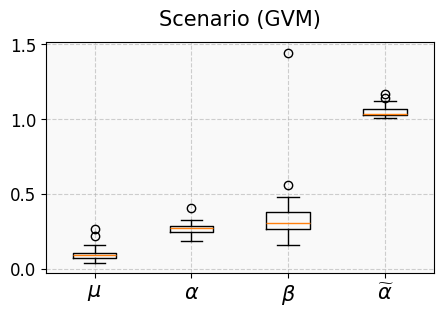}
    \end{subfigure}
    \hspace{0.02\textwidth}
    \begin{subfigure}[t]{0.27\textwidth}
        \centering
        \includegraphics[width=\linewidth]{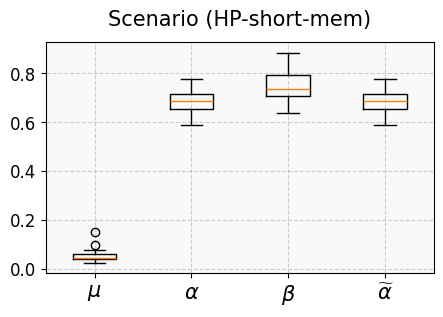}
    \end{subfigure}
    \captionsetup{justification=centering}
    \caption{Boxplots of the relative squared error for each group of parameters ($ \mu,\,  \alpha,\,  \beta$ and $ {\widetilde\alpha}$) under (GVM) for 25 realisations of Scenarios~(HP), (VM), (GVM) and (HP-short-mem), after Step~5 of the estimation procedure.}
    \label{fig:errors_dim10}
\end{figure}

For each scenario, we display the outcomes from Tests~1 to 3, \textcolor{black}{performed using Student-based confidence intervals}, and the relative squared errors for each group of parameters ($ \mu,\,  \alpha,\,  \beta$ and $ {\widetilde\alpha}$) after Step~5, by considering vector norms (unlike the bivariate case, errors after Steps~1 and 3 are omitted).
As shown in Figure~\ref{fig:tests_dim10}, Model~(GVM) accurately recovers parameters and interaction structure in Scenarios~(HP), (VM) and (GVM), with Tests~1 to 3 producing reliable results, and Figure~\ref{fig:errors_dim10} confirming low relative errors after Step~5.
In Scenario~(HP-short-mem), however, relative errors for $\widetilde\alpha$ are high, and Figure~\ref{fig:tests_dim10} shows that Test~2 tends to set a lot of values of $\widetilde\alpha$ to zero.
This is expected, as large values of $\beta$ cause the exponential kernel to decay rapidly, causing the processes to quickly lose memory of past events, even in Model~(HP), leading Model~(GVM) to overestimate values of $\alpha$ and assign zero values to $\widetilde \alpha$.

In Table~\ref{table:pvalues_dim10}, we also display the average $p$-value from the outcomes of the goodness-of-fit test for Model (GVM), by applying the resampling procedure \textcolor{black}{50} times.
These tests are performed with the \textcolor{black}{final estimation obtained after Step~5 over all realisations.}
In comparison, \(p\)-values are also provided for estimations under Models (HP) and (VM), in which case the final estimates are taken after Step~3.
As before, Model~(GVM) yields consistently high $p$-values across all scenarios, while Models (HP) and (VM) only perform reliably when the data match their model.
In Scenario~(HP-short-mem), $p$-values are comparable across models, indicating no clear distinction, even when the model is misspecified.

\begin{table}[ht!]
\centering
\resizebox{\textwidth}{!}{
\begin{tabular}{|c|ccc|ccc|ccc|ccc|}
\hline
Scenario & \multicolumn{3}{c|}{(HP)} & \multicolumn{3}{c|}{ (VM)} & \multicolumn{3}{c|}{ (GVM)} & \multicolumn{3}{c|}{ (HP-short-mem)} \\
\hline
Model & (HP) & (VM) & (GVM) & (HP) & (VM) & (GVM) & (HP) & (VM) & (GVM)  & (HP) & (VM) & (GVM)\\
\hline
MLE  & \textcolor{black}{0.57} & \textcolor{black}{0.3} & \textcolor{black}{0.53} & \textcolor{black}{0.34} & \textcolor{black}{0.45} & \textcolor{black}{0.54}  & \textcolor{black}{0.39} & \textcolor{black}{0.32} & \textcolor{black}{0.49} & \textcolor{black}{0.16} &\textcolor{black}{0.15} & \textcolor{black}{0.26} \\
\hline
\end{tabular}}
\captionsetup{justification=centering}
\caption{Average $p$-values over 25 estimations of realisations from Scenarios~(HP), (VM), (GVM) and (HP-short-mem).}  \label{table:pvalues_dim10}
\end{table}


\subsection{Summary of the numerical study}

Overall, estimation under Model~(GVM) performs consistently well across all models and scenarios, producing satisfactory test results, low relative errors after Step~5, and high $p$-values.
In contrast, Models (HP) and (VM) yield reliable results only when the data is simulated using the same model: when applied to data generated by a different model, these two models exhibit poor performance, particularly in terms of $p$-values.
However, these discrepancies are most pronounced for small values of $\beta$, where the differences between the three models are more distinct.
For larger values of $\beta$, the rapid decay of the exponential distribution causes all three models to quickly lose memory of past events, reducing the impact of model differences.


\section{Application on neuronal data} \label{sect:application}

\subsection{Data preprocessing} \label{sect:preprocessing}

In this section, we apply our estimation procedure under Model~(GVM) to a collection of 10 trials consisting in the measurement of spike trains of 250 neurons from the lumbar spinal circuits of a red-eared turtle \cite{petersen_lognormal_2016,radosevic_decoupling_2019}.
Each trial events were recorded for 40 seconds.
We begin with a pre-processing approach inspired by \cite{bonnet_neuronal_2022}, detailed in Steps~1 to 3 below, which involve trimming the trials to ensure stationarity, removing trials with too many non-active neurons, and retaining only sufficiently active neurons.
After Step~3, the dataset is reduced to observations within $[0,10]$, a total of 7 trials, and 61 neurons.
Due to the limited number of trials, \textcolor{black}{and in order to perform tests in our estimation procedure}, we then follow a resampling approach described in \cite{bonnet_inference_2023} and \cite{reynaud-bouret_goodnessoffit_2014}, corresponding to Steps~4 to 6, \textcolor{black}{in order to generate a larger number of trials}.
The full procedure is described below. 
The choices made at each step are further motivated by the visualisations provided in \ref{app:preprocessing}, which illustrate the normalised cumulative spike counts $N^i(t)/t$, for the raw, filtered, and resampled data, corresponding respectively to the outcomes of Steps~1, 3, and 6.

\begin{center}
\fbox{\begin{minipage}{\textwidth}
\textbf{Data preprocessing steps:}
\begin{enumerate}
    \item \textbf{Step~1} (Trimming): Data is processed in order to take into account stationarity by considering only events that took place on $[11,21]$.
Consequently, each trial is considered within the normalised time window $[0,10]$.
    \item \textbf{Step~2} (Non-relevant trials removal): Only trials with fewer than 10 non-active neurons are retained.
    \item \textbf{Step~3} (Non-active neurons removal): Only neurons with at least 50 jumps across all trials are kept.
    \item \textbf{Step~4} (Subsampling): Three realisations, denoted $(N_1,N_2,N_3)$, are sampled at random without replacement, preserving their order.
    \item \textbf{Step~5} (Concatenation): The three realisations are concatenated by considering that $N_1$ takes place in $[0,10]$, $N_2$ takes place in $[10,20]$ and $N_3$ takes place in $[20,30]$.
    \item \textbf{Step~6} (Resampling): Steps 4 and 5 are repeated 25 times to obtain a sample of 25 realisations.
\end{enumerate}
\end{minipage}}
\end{center}


\subsection{Estimation results}

\begin{figure}[ht!]
\centering
\includegraphics[width=\linewidth]{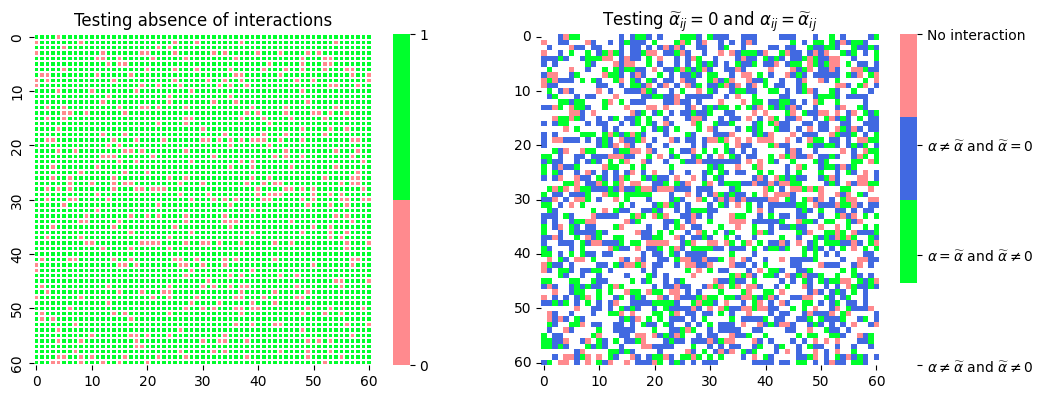}
 \captionsetup{justification=centering}
\caption{Test results from GVM estimations for 25 resampled trials.
On the left: red (0) indicates no interaction, while green (1) indicates a detected interaction.
On the right: summary of the results for Tests~2 and 3.} \label{fig:data_tests}
\end{figure}

\begin{figure}[ht!]
\centering
\includegraphics[width=\linewidth]{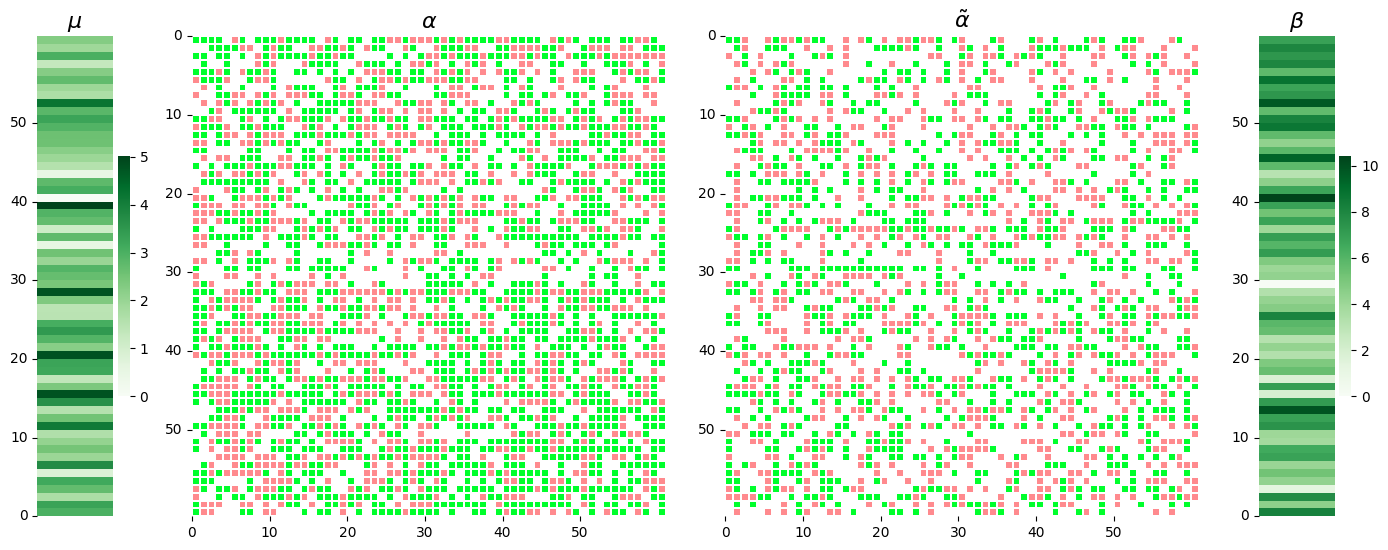}
 \captionsetup{justification=centering}
\caption{Heatmaps of estimated parameters after Step~5.
For $\alpha$ and $\widetilde\alpha$: white represents no interaction, green represents an excitatory interaction, and red represents an inhibitory interaction.} \label{fig:data_heatmaps}
\end{figure}

\textcolor{black}{We follow Steps~1 to 5 of the estimation procedure under Model~(GVM) as described in Section~\ref{sect:tests}.
Yet, since only 7 trials are available but Tests~1, 2 and 3 require a larger number of trials, interaction detection is performed on the 25 resampled trials obtained after Step~6 of the preprocessing described in Section~\ref{sect:preprocessing}, whilst the final estimation is done on the 7 original trials, obtained by maximising the sum of the log-likelihoods.} 
The results of Tests~1 (on the left), 2, and 3 (on the right) are summarised in Figure~\ref{fig:data_tests}.
The results of Test~1 indicate that many significant interactions persist (\textcolor{black}{3264} out of 3721), which can be attributed to our preprocessing procedure that retains only sufficiently active neurons.
On the diagonal values, our test detects 56 interactions out of 61, highlighting the major effect of self-interactions.
The results from Tests~2 and 3 demonstrate that both classic memory dynamics and variable memory dynamics are detected.
In particular, we detect \textcolor{black}{1045} interactions where $\widetilde\alpha_{ij}=0$ and $\alpha_{ij}\neq \widetilde\alpha_{ij}$, and \textcolor{black}{630} interactions where $\alpha_{ij}=\widetilde\alpha_{ij}$ and $\widetilde\alpha_{ij}\neq0$, supporting the relevance of our model and its ability to capture different types of dynamics.
Furthermore, this finding aligns with our model’s framework, where some \textcolor{black}{interactions have a memory reset} while others do not.
Estimation results for final estimations $\mu,\alpha,\beta$ and $\widetilde\alpha$ after Step~5 of the estimation proecedure are presented in Figure~\ref{fig:data_heatmaps}.
The heatmap matrices of the signs of $\alpha$ and $\widetilde\alpha$ show numerous significant interactions, both excitatory and inhibitory.

We also display in Table~\ref{table:pvalues_data} the average $p$-value from the outcomes of the goodness-of-fit test by applying the resampling procedure from Section~\ref{sect:goodness_fit} a total of 50 times \textcolor{black}{for the final estimator}, under Model~(GVM), Model~(HP) \textcolor{black}{and Model~(VM)}.
While the $p$-values for Models~(HP) and (VM) fall below the common rejection thresholds, the $p$-value for Model~(GVM) is higher at \textcolor{black}{0.051}, suggesting it is better suited to the data.

\begin{table}[ht!]
\centering
\begin{tabular}{|c|c|c|c|}
\hline
Model & (HP) & \textcolor{black}{(VM)} & (GVM) \\
\hline
$p$-value  &  \textcolor{black}{$0.0004$} & \textcolor{black}{$0.01$} & \textcolor{black}{$0.051$} \\
\hline
\end{tabular}
\captionsetup{justification=centering}
\caption{Average $p$-values for final estimations over 50 resampled trials.}  \label{table:pvalues_data}
\end{table}


\section{Conclusion and Perspectives} \label{sect:conclusion}

In this paper, we introduced a generalised model for nonlinear Hawkes processes in a finite multidimensional setting, taking into account possible variable memory dynamics.
We proved existence of processes in this framework and provided in particular a weak and nontrivial condition for the variable length memory model.
The question of stationarity remains open and would be a valuable direction for future work, especially in order to derive long-term properties and to justify our statistical procedures.

On the inference side, we
derived explicit formulas for the conditional intensities and developed a methodology for parameter estimation, interaction detection, and distinguishing between classic and variable memory dynamics.
Our numerical results on synthetic data suggest the relevance and the flexibility of this generalised model, and its ability to adapt to both variable and fixed memory scenarios.

We also applied our inference procedure to neuronal spike data, which revealed several cases of variable memory dynamics, \textcolor{black}{and yielding higher $p$-values under our generalised model than for the two pre-existing ones.}
However, \textcolor{black}{Model (GVM) can be refined in order to better suit neuronal data.}
In particular, this work focuses on two main simplifying assumptions: simple exponential kernels and decay rates depending only on the receiving process.
Future research could aim at extending the model to other types of kernels and more general interaction structures.
Besides, neurons exhibit a refractory period following a spike, during which they cannot fire again \cite{galves_probabilistic_2024}.
It would therefore be relevant to account for this biological feature in the model to better distinguish between true memory resets and the effects of refractoriness, for instance by considering that the self-interaction functions are sums of two exponential functions, or are non-parametric functions, as it is done in \cite{bonnet_nonparametric_2025} for classical Hawkes processes without variable memory.

Additionally, according to Dale's principle \cite{strata_dales_1999, galves_probabilistic_2024}, neurons are generally assumed to exert either an excitatory or inhibitory effect on other neurons, so it would be of interest to integrate statistical tests to determine these interaction types.

Finally, our testing procedures rely on the asymptotic normality of the estimators.
Establishing consistency results, as has been done for classic self-exciting Hawkes processes \cite{ogata_asymptotic_1978, lotz_sparsity_2024}, would be an important next step.
A natural starting point would be to derive theoretical guarantees for Model~(HP) with inhibition, or for Model~(GVM) restricted to self-exciting interactions with positive kernels.
In both cases, computing the expected number of events in a finite interval is already nontrivial, making the extension of existing consistency results particularly challenging.

\section*{Acknowledgements}
    \textcolor{black}{The three authors would like to thank the anonymous reviewers for their constructive comments, which helped improve the estimation procedure and the manuscript.}
    They would also like to warmly thank Eva Löcherbach for insightful suggestions regarding variable length memory models for neuroscience,
    and Arnaud Guyader for fruitful discussions on continuous Markov chains.
   This work is part of the project HAPPY ANR-23-CE40-0007 and  2022 DAE 103 EMERGENCE(S) - PROCECO.




\appendix

\section{Proof of Proposition~\ref{prop:existence GVM}} \label{app:existence GVM}
	Let us keep notations from Section~\ref{sect:def VM hawkes}.
Denote $(T_k^j)_{k\in \N^*}$ the arrival times for each process $N^j$.
Then, by monotony and sublinearity of $x\mapsto x^+$ on $\R$, for any $1\leq i\leq d$, for any $t\geq 0$, a.s.:
	\begin{align*}
		\lambda^i(t) & = \left(\mu_i + \sum_{j=1}^d\left[ \sum\limits_{k\in \N^*,\, T_{N^i(t)}^i\leq T_k^j <t} h_{ij}(t-T_k^j) + \sum\limits_{k\in \N^*,\, T_k^j < T_{N^i(t)}^i} \widetilde h_{ij}(t-T_k^j)\right] \right)^+ \\
		& \leq \left(\mu_i + \sum_{j=1}^d\left[ \sum\limits_{k\in \N^*,\, T_{N^i(t)}^i\leq T_k^j <t} \max(h_{ij},\widetilde h_{ij})(t-T_k^j) + \sum\limits_{k\in \N^*,\, T_k^j < T_{N^i(t)}^i} \max(h_{ij},\widetilde h_{ij})(t-T_k^j)\right] \right)^+\\
		& = \left(\mu_i + \sum_{j=1}^d\sum_{T_k^j < t} \max(h_{ij},\widetilde h_{ij})(t-T_k^j) \right)^+ \leq \mu_i + \sum_{j=1}^d \sum_{T_k^j < t} \max(h_{ij},\widetilde h_{ij})^+(t-T_k^j)
	\end{align*}
thus, a.s.:, $\sum_{i=1}^d \lambda^i(t) \leq \sum_{i=1}^d \overline \lambda^i(t)$, where
$$\overline \lambda^i(t) =  \mu_i + \sum_{j=1}^d \sum_{T_k^j < t} \max(h_{ij},\widetilde h_{ij})^+(t-T_k^j)\,.$$

By the spectral radius assumption and by \cite[Theorem 7]{bremaud_stability_1996}, there exists a family $(N^1,...,N^d)$ of point processes with respective conditional intensities $\overline\lambda^{1},...,\overline\lambda^{d}$.
By locally upper bounding the sum $\sum_{i=1}^d \overline\lambda^{d}$ between two consecutive jump times, we can construct a one-dimensional pathwise constant $(\mathcal F_t^-)_{t\geq 0}$-predictable process $\overline\lambda(t)$ (where $(\mathcal F_t)_{t\geq 0}$ is the canonical filtration of $N$), and a one-dimensional stationary point process $\overline N$ with conditional intensity $\overline \lambda$, such that $\sum_{i=1}^d \lambda^i(t) \leq \overline\lambda(t)$ a.s. Thus, by thinning according to \cite[Proposition 1]{ogata_lewis_1981}, there exists a multivariate point process $N=(N^1,...,N^d)$ with conditional intensities $\lambda^1,...,\lambda^d$.


\section{Proof of Theorem~\ref{thm:existence VM}} \label{app:existence VM}

For any $1\leq i \leq d$, denote $(T_k^i)_{k\in \N^*}$ the arrival times for $N^i$.
Then, for any $t\geq 0$:
\begin{align*}
	\lambda^i(t) & = \left( \mu_i + h_{ii}(t-T^i_{N^i(t)}) + \sum_{j\neq i} \sum_{T^i_{N^i(t)} \leq T_k^j < t} h_{ij}(t-T_k^j) \right)^+ \\
	& \leq \mu_i + \left|h_{ii}(t-T^i_{N^i(t)})\right| + \sum_{j\neq i} \sum_{T^i_{N^i(t)} \leq T_k^j < t} \left|h_{ij}(t-T_k^j)\right| \\
	& \leq C \left( 1+ \sum_{j\neq i}  N^j\left[T_{N^i(t)}^i,t\right)\right) =: \overline \lambda^i(t) ,
\end{align*}
where $C := \underset{1\leq i,j\leq d}{\sup} (\mu_i + \|h_{ij}\|_\infty).$
Therefore, we can dominate the original intensity $\lambda=(\lambda^1,\dots,\lambda^d)$ by a piecewise constant stochastic process $\overline\lambda=(\overline \lambda^1,\dots,\overline \lambda^d)$, constant between two consecutive event times.
By \cite[Proposition 1]{ogata_lewis_1981}, the Hawkes process $N$ can then be constructed via thinning from a dominating process $\overline N$ with intensity $\overline \lambda$.
In particular:
$$\forall A\in \mathcal B(\R^d),\, \overline N(A)<\infty \implies N(A)<\infty\,.$$

We now focus on the process $\overline N = (\overline N^1, \dots, \overline N^d)$, defined by its intensity $\overline\lambda$.
Our goal is to prove that $\overline N$ does not explode (which implies that the process $N$ does not explode).
To do so, we will first show that the intensity process $\overline\lambda$ forms a time-continuous Markov chain.
This structure then allows us to apply results from Markov process theory to establish the non-explosion of $\overline N$.
The intensity $\overline\lambda$ takes values in:
$$E = \bigcup_{i=1}^d  \{ (C (1+k_1),\dots,C, \dots, C(1+ k_d)), \, (k_1,\dots,k_d)\in \N^d \}\, .$$
Here, the entry $C$ at position $i$ indicates that process $i$ has just experienced a jump.
The values $C(1 + k_j)$ represent the self-excitation effect (the "+1") combined with the cumulative contribution from the $d - 1$ other processes since the last event in dimension $j$.

For any $l\in \N$, denote $Z^{(l)}\in E$ the value of the intensity $\overline \lambda$ on the interval $(\overline T_{(l)},\overline T_{(l+1)})$:
$$Z^{(l)} = \left(\overline \lambda^1(T_{(l)}^+),\dots, \overline \lambda^d(T_{(l)}^+)\right)\,.$$

Initially, we have $Z^{(0)}=(C,\dots,C)$.
Now suppose that $Z^{(l)}= ( Z^{(l)}_1,\dots,Z^{(l)}_d)$.
If the next event time is from dimension $j\in \{1,\dots,d\}$, which occurs with probability: $$Z_j^{(l)}/\sum_{k=1}^d Z_k^{(l)}\,,$$
the next state is given by:
$$Z^{(l+1)}= (Z^{(l)}_1+C,\dots,C,\dots, Z^{(l)}_d+C)\,,$$

where the value $C$ appears at position $j$ (corresponding to the dimension that just jumped), while the others are incremented by $C$.
This defines a Markov chain $(Z^{(l)})_{l\in \N^*}$ taking values in $E$, with transition probabilities $(p_{zz'})_{z,z'\in E}$ as described above.
Thus, the process $(\overline \lambda(t))_{t\geq 0}$ is a time-continuous Markov chain with discrete state space.
The chain starts at $z_0:=(C,\dots,C)$, which is visited only once at time 0.
After the first jump, the chain evolves within an irreducible class $E' = E\backslash\{z_0\}$.
Recurrence of the chain can thus be studied by restricting to $E'$.
According to \cite[Proposition 2.4 (p.44)]{asmussen_applied_2003}, a sufficient condition to ensure non-explosion of the process is that this restricted chain is recurrent.

The remainder of the proof focuses on establishing the recurrence of the chain.
We treat the cases $d=2$ and $d>2$ separately: for $d=2$, the transition graph is simple, making the proof more intuitive.
For $d>2$, the graph becomes significantly more complex, and the proof relies on a recurrence criterion.

\subsection*{Proof of recursion for $d=2$} 

If $d=2$, the state space is:
\begin{align*}
    E := & \{(C,  C(1 + k)),\, k\in \N \} \cup  \{(C(1+k), C), \, k\in \N\} \\
    & =: \{z^{(1)}_k,\, k\in \N\} \cup \{z_k^{(2)},\, k\in \N\}\,,
\end{align*}

and the transition graph of the chain $(Z^{(l)})_{l\in \N}$ is illustrated below, with transition probabilities $(p_{z,z'})_{z,z'\in E}$ as described above.
Intuitively, the chain's dynamics are governed by the occurrence of events in the two underlying processes: when events are generated by the same process, the chain follows one of the two possible branches and continues to evolve in that direction.
When an event is triggered by the other process, the chain returns to the state corresponding to the other branch.

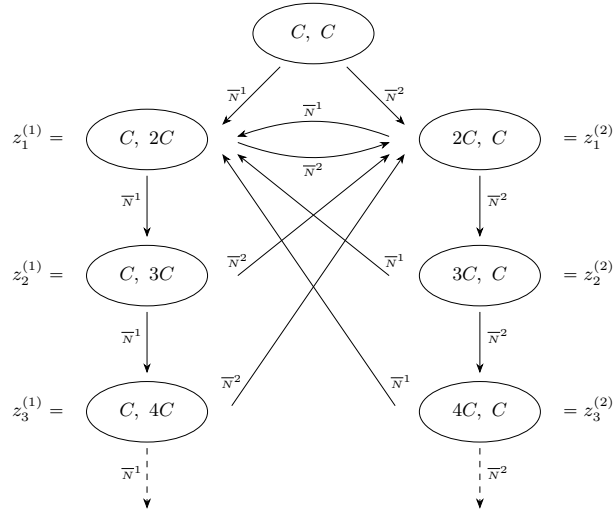
\begin{figure}[ht!]
\centering
\resizebox{0.55\textwidth}{!}{%
\begin{circuitikz}
\tikzstyle{every node}=[font=\tiny]

\draw  (13.25,12.75) ellipse (1cm and 0.5cm);
\node [font=\small] at (13.25,12.75) {$C,\,\, C$};

\draw [line width=0.2pt, ->, >=Stealth] (13.8,12.2) -- (14.75,11.25);
\draw [line width=0.2pt, ->, >=Stealth] (12.7,12.2) -- (11.75,11.25);

\node [font=\tiny] at (12,11.75) {$\overline N^1$};
\node [font=\tiny] at (14.6,11.75) {$\overline N^2$};

\draw  (16,11) ellipse (1cm and 0.5cm);
\node [font=\small] at (16,11) {$2C,\,\, C$};
\node [font=\small] at (17.8,11.05) {$=z^{(2)}_1$};

\draw  (10.5,11) ellipse (1cm and 0.5cm);
\node [font=\small] at (10.5,11) {$C,\,\, 2C$};
\node [font=\small] at (8.7,11.05) {$z^{(1)}_1=$};

\draw [line width=0.2pt, ->, >=Stealth] (12,10.95) to [bend right = 20] (14.5,10.95);
\draw [line width=0.2pt, ->, >=Stealth] (14.5,11.05) to [bend right = 20] (12,11.05);

\node [font=\tiny] at (13.25,11.5) {$\overline N^1$};
\node [font=\tiny] at (13.25,10.5) {$\overline N^2$};


\draw  (16,8.75) ellipse (1cm and 0.5cm);
\node [font=\small] at (16,8.75) {$3C,\,\, C$};
\node [font=\small] at (17.8,8.8) {$=z^{(2)}_2$};

\draw  (16,6.5) ellipse (1cm and 0.5cm);
\node [font=\small] at (16,6.5) {$4C,\,\, C$};
\node [font=\small] at (17.8,6.55) {$=z^{(2)}_3$};

\draw  (10.5,8.75) ellipse (1cm and 0.5cm);
\node [font=\small] at (10.5,8.75) {$C,\, \,3C$};
\node [font=\small] at (8.7,8.8) {$z^{(1)}_2=$};

\draw  (10.5,6.5) ellipse (1cm and 0.5cm);
\node [font=\small] at (10.5,6.5) {$C,\,\, 4C$};
\node [font=\small] at (8.7,6.55) {$z^{(1)}_3=$};


\draw [line width=0.2pt, ->, >=Stealth] (16,10.4) -- (16,9.4);
\draw [line width=0.2pt, ->, >=Stealth] (16,8.15) -- (16,7.15);

\draw [line width=0.2pt, ->, >=Stealth] (10.5,10.4) -- (10.5,9.4);
\draw [line width=0.2pt, ->, >=Stealth] (10.5,8.15) -- (10.5,7.15);


\draw [line width=0.2pt, ->, >=Stealth] (12,8.75) -- (14.5,10.75);
\draw [line width=0.2pt, ->, >=Stealth] (11.9,6.6) -- (14.75,10.75);

\draw [line width=0.2pt, ->, >=Stealth] (14.5,8.75) -- (12,10.75);
\draw [line width=0.2pt, ->, >=Stealth] (14.6,6.6) -- (11.75,10.75);


\draw [->, >=Stealth, dashed] (10.5,5.9) -- (10.5,4.9);
\draw [->, >=Stealth, dashed] (16,5.9) -- (16,4.9);

\node [font=\tiny] at (10.25,10) {$\overline N^1$};
\node [font=\tiny] at (10.25,7.75) {$\overline N^1$};
\node [font=\tiny] at (10.25,5.5) {$\overline N^1$};

\node [font=\tiny] at (16.3,10) {$\overline N^2$};
\node [font=\tiny] at (16.3,5.5) {$\overline N^2$};
\node [font=\tiny] at (16.3,7.75) {$\overline N^2$};

\node [font=\tiny] at (11.9,7) {$\overline N^2$};
\node [font=\tiny] at (12,9) {$\overline N^2$};

\node [font=\tiny] at (14.7,7) {$\overline N^1$};
\node [font=\tiny] at (14.6,9) {$\overline N^1$};

\end{circuitikz}
}%
 \captionsetup{justification=centering}
\caption{Transition graph of the Markov chain $(Z^{(l)})_{l\in \N}$ for $d=2$.} 
\label{fig:my_label}
\end{figure}

Let us show that the chain restricted to $E\backslash\{z_0\}$ is recurrent.
Since this chain is irreducible, it suffices to show that the state $z_1^{(1)}$ is recurrent.
We define the return time to $z_1^{(1)}$:
$$T_{z_1^{(1)}} = \text{ inf } \{ n\in \N^*,\, Z^{(n)} = z_1^{(1)} \, | Z^{(0)} = z_1^{(1)}\}\,.$$
Denote $\P_1 := \P_{z_1^{(1)}}$.
Our goal is to show that $\P_1(T_{z_1^{(1)}} = \infty) =0$, meaning that the chain returns to $z_1^{(1)}$ with probability 1.
Examining the graph of the Markov chain, for $T_{z_1^{(1)}} $ to be infinite, one of two scenarios must occur: either only event times from $\overline N^1$ are observed indefinitely, or at some point, an event time from $\overline N^2$ occurs, after which only event times from $\overline N^2$ are observed. 
Therefore, we can decompose the probability as follows:
\begin{align*}
	\P_1(T_{z_1^{(1)}} = \infty) & = \P_1 \left( \forall l\geq 1,\, \overline d_l = 1\right) + \P_1\left(  \exists k\geq 1,\, \forall l\leq k,\,  \overline d_l= 1, \text{  and } \forall l>k,\, \overline d_l = 2\right) \\
	& \leq \P_1\left( \forall l\geq 1,\, \overline d_l = 1\right) + \sum_{k\geq 1} \P_1\left(\forall l>k,\, \overline d_l = 2 \right).
\end{align*}
However, $\P_1 \left( \forall l\geq 1,\, \overline d_l = 1\right) = \underset{p\rightarrow \infty}{\text{ lim }} \P_1 \left( \forall 1\leq l\leq p,\, \overline d_l = 1\right)$, with:
\begin{align*}
	\P_1 \left( \forall 1\leq l \leq p,\, \overline d_l = 1\right) & =  p_{z_1^{(1)}, z_2^{(1)}} \dots p_{z_{p-1}^{(1)}, z_p^{(1)}}  = \prod_{l=1}^{p-1} \textcolor{black}{\frac{C}{C+lC}}\\
	& \textcolor{black}{=\frac{1}{p!}} \underset{p\rightarrow \infty}{\longrightarrow} 0.
\end{align*}
Hence, we have $\P_1\left( \forall l\geq 1,\, \overline d_l = 1\right) =0$.
Similarly, for any $k\geq 1$, $ \P_1\left(\forall l>k,\, \overline d_l = 2 \right) =0$.
It follows that $\P_1(T_{z_1^{(1)}} = \infty) =0$, i.e.
$z_1^{(1)}$ is recurrent.

\subsection*{Proof of recursion for $d>2$} 

According to \cite[Proposition 5.3]{asmussen_applied_2003}, the chain is recurrent if there exists a finite subset $E_0$ of $E$ and a function $g:E\rightarrow\R$ such that $g(z)\underset{\|z\|\rightarrow \infty}{\longrightarrow} +\infty$ and:
$$\forall z \notin E_0,\, \sum_{z'\in E} p_{zz'} g(z') \leq g(z)\,.$$
Let us consider the following set $E_0$ and the following function $g: E\rightarrow \R$:
$$E_0= \{z\in E,\, \|z\|_1\leq C d^2\}\,, g(z)=\sum_{i=1}^d z_i\,.$$
Then $E_0$ is finite and $g(z)\underset{\|z\|\rightarrow \infty}{\longrightarrow} +\infty$.
Let $z=(z_1,\dots,z_d)\notin E_0$.
Then the future possible states for the Markov chain are $z^i := (z_1+C,\dots,C,\dots,z_d+C)$ for any $1\leq i \leq d$.
Therefore:
\begin{align*}
	\sum_{z'\in E} p_{zz'}g(z') & = \sum_{i=1}^d \frac{z_i}{g(z)} g(z^i)  =\sum_{i=1}^d \frac{z_i}{g(z)} \left( \sum_{j\neq i} z_j + dC\right) \\
	& = \sum_{i=1}^d \frac{z_i}{g(z)} \left( g(z) - z_i +dC \right) = g(z) + \sum_{i=1}^d \frac{z_i}{g(z)}  \left( - z_i + dC\right) \\
	& = g(z) - \frac{\|z\|_2^2}{g(z)} + dC \sum_{i=1}^d \frac{z_i}{g(z)}  \leq g(z) - \frac{\|z\|_2^2}{g(z)} +dC\,.
\end{align*}
However:
$$\frac{\|z\|_2^2}{g(z)} = \frac{\|z\|_2^2}{\|z\|_1} \geq \frac{\|z\|_1}{d} > dC,$$
which implies:
$$\sum_{z'\in E} p_{zz'}g(z') \leq g(z)\,,$$
thus the chain is recurrent.


\section{Proof of Proposition~\ref{prop:recursion multivariate VM hawkes} } \label{app:recursion multivariate VM hawkes}

We keep the same notations: we denote $(T_k^j)_{k\in \N^*}$ the arrival times for each process $N^j$, and $(T_{(l)})_{l\in \N^*}$ the ordered sequence of arrival times of the process $N=(N^1,...,N^d)$ with $d_l$ being the dimension which the arrival time $T_{(l)}$ belongs to.
Let $l\in \N^*$.
We separate cases $t\in ]T_{(l)},T_{(l+1)}[$ and $t=T_{(l+1)}$.	
If $t\in ]T_{(l)},T_{(l+1)}[$: we start by noting that, for any $1\leq i \leq d$: $T^i_{N^i(t)}=T^i_{N^i(T_{(l)})}\,,$ therefore:
	\begin{align*}
		\eta^i(t) & = \sum_{j=1}^d \sum_{T_{N^i(t)}^i\leq T_k^j <t} \alpha_{ij} e^{-\beta_i(t-T_k^j) }= e^{-\beta_i (t-T_{(l)})} \sum_{j=1}^d \sum_{T_{N^i(T_{(l)})}^i\leq T_k^j <t} \alpha_{ij} e^{-\beta_i(T_{(l)}-T_k^j) }\,, 
	\end{align*}
	
	and, by the assumption that two different processes cannot jump at the same time:
$$\sum\limits_{T_{N^i(T_{(l)})}^i\leq T_k^j <t} \alpha_{ij} e^{-\beta_i(T_{(l)}-T_k^j) } = \left\{
	\begin{array}{ll}
		\sum\limits_{T_{N^i(T_{(l)})}^i\leq T_k^j <T_{(l)}} \alpha_{ij} e^{-\beta_i(T_{(l)}-T_k^j) } & \text{ if } j\neq d_l\\
		\sum\limits_{T_{N^i(T_{(l)})}^i\leq T_k^j <T_{(l)}} \alpha_{ij} e^{-\beta_i(T_{(l)}-T_k^j) } +\alpha_{i,d_l} & \text{ if } j=d_l
		\end{array}
	\right.,$$
	
	therefore:
	$$\eta ^i(t) = (\eta^i(T_{(l)}) + \alpha_{i,d_l})e^{-\beta_i(t-T_{(l)})}\,.
$$
	
	Similarly, we also have: $\widetilde\eta_{aux}^i(t) = (\widetilde\eta_{aux}^i(T_{(l)}) + \widetilde\alpha_{i,d_l})e^{-\widetilde\beta_i(t-T_{(l)})}.$
	For $\widetilde \eta^i$, we have:
	\begin{align*}
		\widetilde \eta^i(t) & = \sum_{j=1}^d \sum_{T_k^j < T_{N^i(t)}^i} \widetilde\alpha_{ij} e^{-\widetilde\beta_i(t-T_k^j) } = e^{-\widetilde\beta_i (t-T_{(l)})} \sum_{j=1}^d \sum_{T_k^j <T_{N^i(T_{(l)})}^i} \widetilde\alpha_{ij} e^{-\widetilde\beta_i(T_{(l)}-T_k^j) } =  e^{-\widetilde\beta_i (t-T_{(l)})}  \widetilde \eta^i(T_{(l)})\,.
	\end{align*}

If $t=T_{(l+1)}$, we have:
	$$T^i_{N^i(T_{(l+1)})} = \left\{ 
	\begin{array}{ll}
		T^i_{N^i(T_{(l)})} & \text{ if } i\neq d_{l+1} \\
		T_{(l+1)} & \text{ if } i=d_{l+1}
	\end{array}
	\right.,$$
	
	therefore if $i\neq d_{l+1}$, then the results for $t\in ]T_{(l)},T_{(l+1)}[$ hold, i.e.:
	\begin{center}
		$\eta^i (T_{(l+1)}) = (\eta^i(T_{(l)}) + \alpha_{i,d_l})e^{-\beta_i(T_{(l+1)}-T_{(l)})},$\\
		$ \widetilde \eta_{aux} ^i(T_{(l+1)}) = (\widetilde \eta_{aux} ^i(T_{(l)}) + \widetilde\alpha_{i,d_l})e^{-\widetilde\beta_i(T_{(l+1)}-T_{(l)})},$ and \\
		$\widetilde \eta^i(T_{(l+1)}) = \widetilde \eta ^i (T_{(l)}) e^{-\widetilde \beta_i (T_{(l+1)}-T_{(l)})}\,.$
	\end{center}

	If $i=d_{l+1}$, then $T^i_{N^i(T_{(l)})} =T_{(l+1)}$, so there are no jumps from any process in $[T^i_{N^i(T_{(l+1)})} ,T_{(l+1)}[$, therefore:
	\begin{center}
		$\eta^i(T_{(l+1)}) = 0$ and $\widetilde\eta_{aux}^i(T_{(l+1)}) = 0\,$,
	\end{center}
	
	 and, for $\widetilde \eta$:
	\begin{align*}
		\widetilde \eta^i(T_{(l+1)})  &=  \sum_{j=1}^d \sum_{T_k^j < T_{N^i(T_{(l+1)})}^i} \widetilde \alpha_{ij} e^{-\widetilde \beta_i (T_{(l+1)}-T_k^j)} \\
		& = e^{-\widetilde \beta_i (T_{(l+1)}-T_{(l)})}  \sum_{j=1}^d \sum_{T_k^j < T_{(l+1)}} \widetilde \alpha_{ij} e^{-\widetilde \beta_i (T_{(l)}-T_k^j)}\\
		& =  e^{-\widetilde \beta_i (T_{(l+1)}-T_{(l)})} \left(  \sum_{j=1}^d \sum_{T_k^j < T^i_{N^i(T_{(l)})}} \widetilde \alpha_{ij} e^{-\widetilde \beta_i (T_{(l)}-T_k^j)} +\sum_{j=1}^d \sum_{T^i_{N^i(T_{(l)})} \leq T_k^j < T_{(l+1)}} \widetilde \alpha_{ij} e^{-\widetilde \beta_i (T_{(l)}-T_k^j)} \right) \\
		& = e^{-\widetilde \beta_i (T_{(l+1)}-T_{(l)})} \left(  \widetilde \eta^i(T_{(l)}) +\sum_{j=1}^d \sum_{T^i_{N^i(T_{(l)})} \leq T_k^j < T_{(l+1)}} \widetilde \alpha_{ij} e^{-\widetilde \beta_i (T_{(l)}-T_k^j)} \right),
	\end{align*}
	
	and:
	\begin{align*}
		\sum_{j=1}^d \sum_{T^i_{N^i(T_{(l)})} \leq T_k^j < T_{(l+1)}} \widetilde \alpha_{ij} e^{-\widetilde \beta_i (T_{(l)}-T_k^j)} & = \sum_{j=1}^d \sum_{T^i_{N^i(T_{(l)})} \leq T_k^j < T_{(l)}} \widetilde \alpha_{ij} e^{-\widetilde \beta_i (T_{(l)}-T_k^j)} + \widetilde \alpha_{i,d_l}\\
		& = \widetilde\eta_{aux}^i(T_{(l)}) + \widetilde\alpha_{i,d_l}\,,
	\end{align*}	
	
	therefore:
	$$\widetilde \eta^i(T_{(l+1)})  = e^{-\widetilde \beta_i (T_{(l+1)}-T_{(l)})} \left( \widetilde \eta ^i(T_{(l)}) + \widetilde\eta_{aux}^i(T_{(l)}) + \widetilde\alpha_{i,d_l}\right).$$


\section{Illustration of Preprocessing Steps} \label{app:preprocessing}

\begin{figure}[ht!]
\centering
\includegraphics[width=0.6\linewidth]{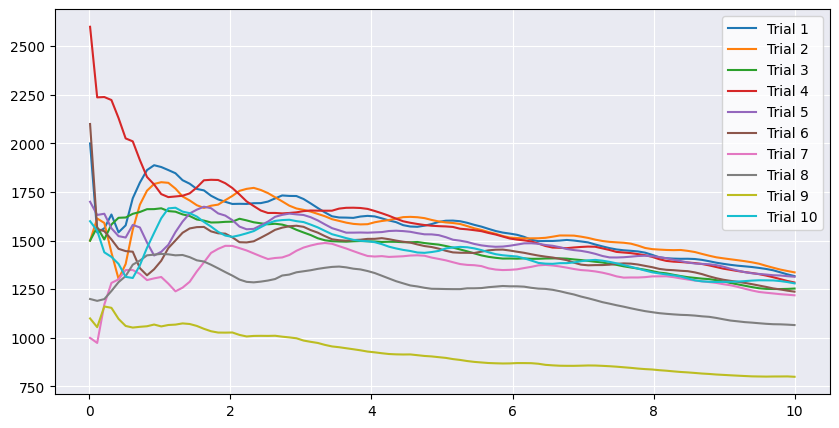}
 \captionsetup{justification=centering}
\caption{Normalised cumulative spike counts \( N^i(t)/t \) for the 10 original trials, shown on the normalised time interval \( [0,10] \), obtained after Step~1.} 
 \label{fig:raw}
\end{figure}

\begin{figure}[ht!]
\centering
\includegraphics[width=0.6\linewidth]{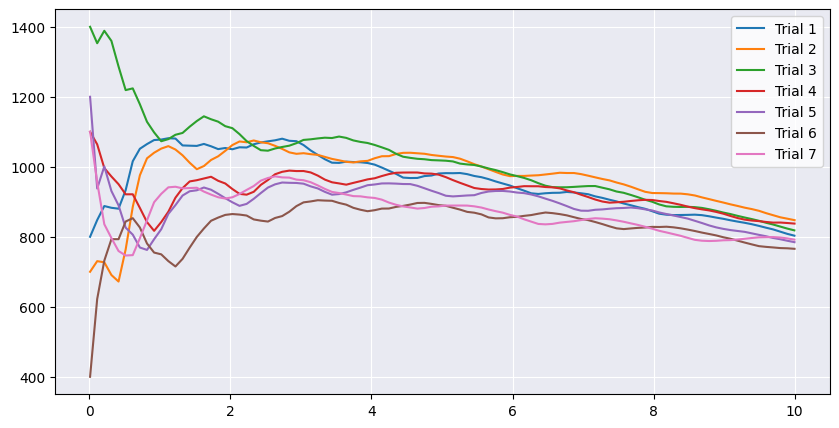}
 \captionsetup{justification=centering}
\caption{Normalised cumulative spike counts \( N^i(t)/t \) for the 7 trials retained after Step~3.} 
\label{fig:filtered}
\end{figure}

\begin{figure}[ht!]
\centering
\includegraphics[width=0.77\linewidth]{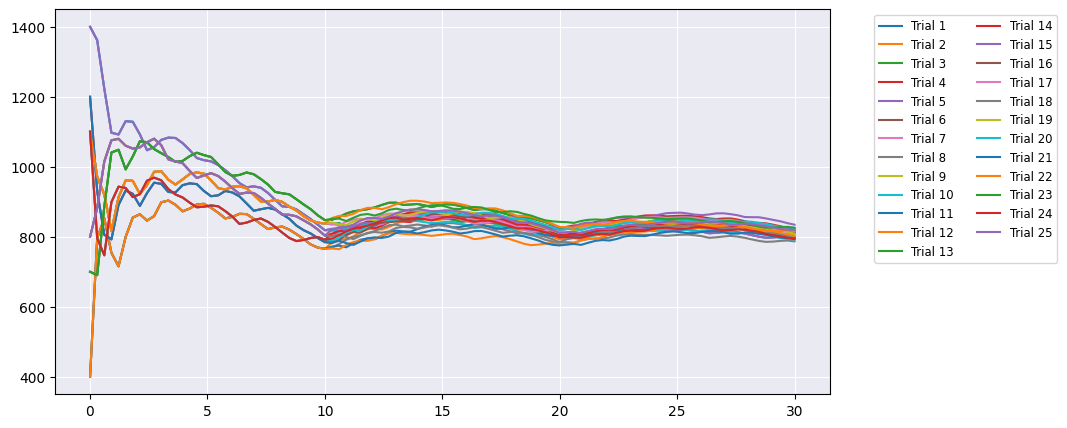}
 \captionsetup{justification=centering}
\caption{Normalised cumulative spike counts \( N^i(t)/t \) for the 25 trials obtained after the resampling procedure in Step~6.} 
\label{fig:resampled}
\end{figure}

Figures~\ref{fig:raw}, \ref{fig:filtered} and \ref{fig:resampled} respectively display the normalised cumulative spike counts \( N^i(t)/t \) for the original trials on the normalised time window $[0,10]$, for the trials after removing non-relevant ones and non-active neurons, and for the 25 resampled trials on the time window $[0,30]$, as explained in Section~\ref{sect:application}. 

\end{document}


{
\title{\textbf{Supplementary material to: Hawkes Processes with Variable Length Memory: Existence, Inference and Application to Neuronal Activity}}
\author{Sacha Quayle\thanks{Sorbonne Universit\'e, CNRS, Laboratoire de Probabilit\'es, Statistique et Mod\'elisation, Paris, France}, Anna Bonnet\footnotemark[1], Maxime Sangnier\footnotemark[1]}
\maketitle
}

\renewcommand{\thesection}{S\arabic{section}}
\renewcommand{\theequation}{S\arabic{equation}}
\renewcommand{\thefigure}{S\arabic{figure}}
\renewcommand{\thetable}{S\arabic{table}}

This supplementary material provides additional numerical experiments and comparisons to justify the methodological choices made in the main text.

\section{Choice of confidence intervals}

\begin{figure}[ht!]
\centering
    \includegraphics[width=0.45\textwidth]{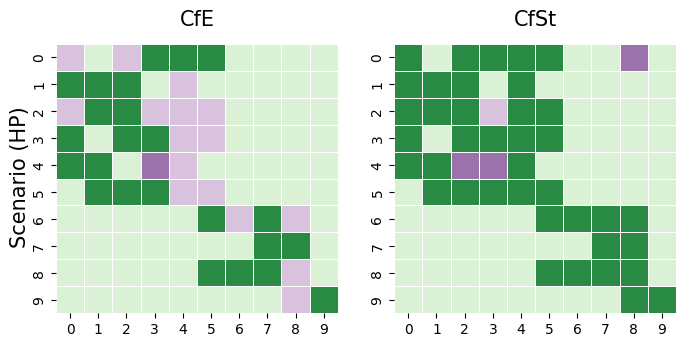}\\[1ex] 
     \includegraphics[width=0.45\textwidth]{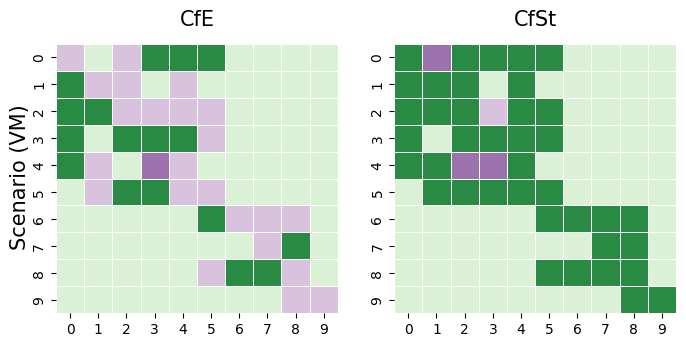}\\[1ex] 
      \includegraphics[width=0.45\textwidth]{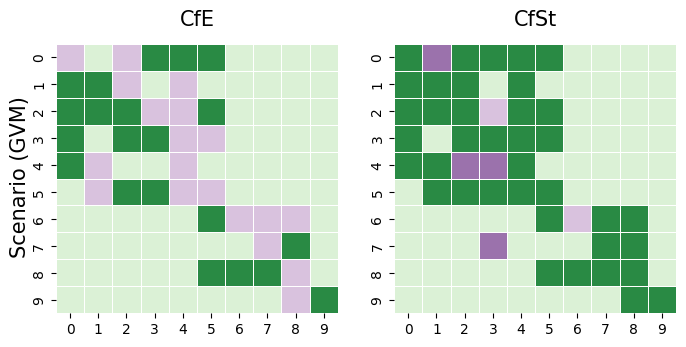}\\[1ex]
 \captionsetup{justification=centering}
\caption{Comparison of both approaches (CfE and CfSt) for detecting absence of interactions across the three scenarios.
Colours mean true positive \textcolor{truezero}{\(\blacksquare\)} (detected 0 or equal value),
true negative \textcolor{truenonzero}{\(\blacksquare\)} (detected non-null value or non-equal value),
false positive \textcolor{falsezero}{\(\blacksquare\)} (non-null value set to 0 or non-equal value set equal),
false negative \textcolor{undetectedzero}{\(\blacksquare\)} (undetected 0 or equal value).} \label{fig:comparison_cf}
\end{figure}

We compare empirical confidence intervals (CfE) with asymptotic normality-based confidence intervals (CfSt). 
We show that the latter leads to fewer errors, and further support this choice by examining the empirical distribution of the estimators. 
We consider the three scenarios (HP), (VM), and (GVM) from the 10-dimensional study presented in the main article.
For each scenario, we simulate 25 independent realisations containing 5000 event times, and estimate the parameters.
Tests for the absence of interactions are then performed, corresponding to Steps~1 and~2 of the estimation procedure. 
In Step~2, we compare the results obtained using CfE and CfSt, as illustrated in Figure~\ref{fig:comparison_cf}, showing that CfSt produces fewer errors.

\begin{figure}[ht!]
    \centering

    \begin{subfigure}[t]{0.48\textwidth}
        \centering
        \includegraphics[width=\linewidth]{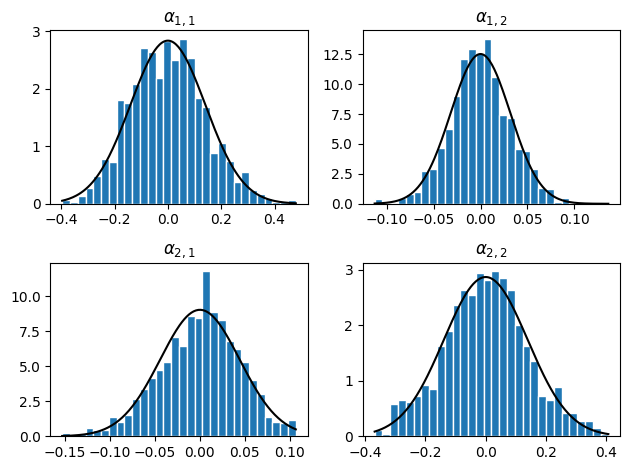}
    \end{subfigure}
    \hspace{0.02\textwidth}
    \begin{subfigure}[t]{0.48\textwidth}
        \centering
        \includegraphics[width=\linewidth]{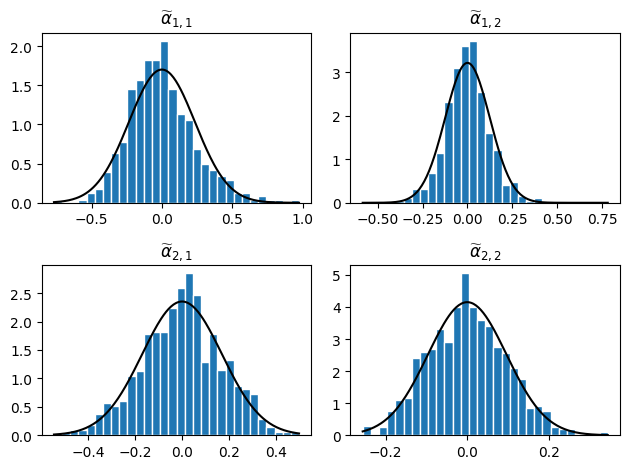}
    \end{subfigure}

    \captionsetup{justification=centering}
    \caption{Centered and normalised histograms of the estimators of $\alpha$ and $\widetilde{\alpha}$, based on 1000 simulated realisations under Scenario (HP).}
   \label{fig:histograms_alpha}
\end{figure}

To further justify the use of asymptotic normality-based confidence intervals, we also examine the empirical distributions of the estimators.
Here, we consider Scenario (HP) from the 2-dimensional study in order to simulate a larger number of realisations. 
We simulate 1000 independent realisations, each containing 5000 event times, and estimate the parameters under Model (GVM). 
We then plot the centered normalised histograms of the estimators for $\alpha$ and $\widetilde{\alpha}$ and the corresponding normal densities, as illustrated in Figure~\ref{fig:histograms_alpha}.

\section{Computation of the final estimator}

\begin{figure}[ht!]
\centering
\includegraphics[width=0.7\linewidth]{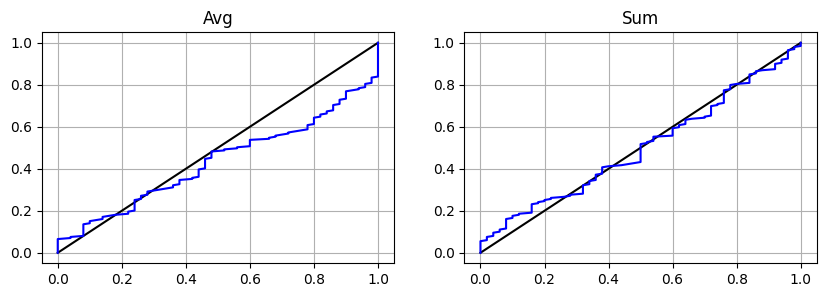}
 \captionsetup{justification=centering}
\caption{Distribution of $p$-values obtained from the goodness-of-fit test over 50 repetitions of the resampling procedure, comparing the two final estimators (Avg and Sum).}  \label{fig:comparison_estimators}
\end{figure}

We compare two approaches for computing the final estimator in Step~5: averaging the estimators obtained from each realisation (Avg), and maximising the sum of log-likelihoods over all realisations (Sum). 
The second method is preferred to obtain a final estimator, as it yields better goodness-of-fit $p$-values and ensures that the resulting parameters define a conditional intensity that remains non-negative at the observed event times.
We consider Scenario (HP) from the 2-dimensional study and perform Steps~1 to~5 of the estimation procedure under Model (GVM). 
In Step~5, the final estimator is computed using both approaches.
Figure~\ref{fig:comparison_estimators} illustrates the distribution of $p$-values obtained after applying the resampling procedure for goodness-of-fit 50 times, while Table~\ref{table:comparison_estimators} reports the average $p$-values and the corresponding log-likelihood values for both estimators, showing that the second estimator (Sum) performs better.

\begin{table}[ht!]
\centering
\begin{tabular}{|c|c|c|}
\hline
Method & Average $p$-value & Log-likelihood \\
\hline
Avg & 0.43 & -53308 \\
\hline
Sum & 0.5 & -53303 \\
\hline
\end{tabular}
\captionsetup{justification=centering}
\caption{Comparison of the two final estimators (Avg and Sum) in terms of average $p$-values from the goodness-of-fit test and corresponding log-likelihood values.}  \label{table:comparison_estimators}
\end{table}

\newpage 
\section{Comparison of goodness-of-fit tests}

In this section, we compare the Kolmogorov–Smirnov (KS) and Cramér–von Mises (CvM) goodness-of-fit tests. 
The CvM test is based on the integrated squared difference between the empirical and theoretical cumulative distribution functions and is generally more powerful than the KS test. 

To further justify this choice, we conduct the following study: we consider Scenario (HP) from the 10-dimensional setting and compute $p$-values under Model (GVM) using the true parameters. 
The resampling procedure is repeated 1000 times, and we evaluate the average $p$-values and the rejection rates (RR) for several significance levels.
The results are reported in Table~\ref{table:comparison_ks_cvm}, showing that the rejection rates for the CvM test are closer to the fixed levels than the KS test.

\begin{table}[ht!]
\centering
\begin{tabular}{|c|c|c|c|c|c|c|}
\hline
Method & Average $p$-value & RR (0.01) & RR (0.025) & RR (0.05) & RR (0.1) & RR (0.2) \\
\hline
KS & 0.48 & 0.015 & 0.034 & 0.064 & 0.12 & 0.22 \\
\hline
CvM & 0.495 & 0.0103 & 0.029 & 0.053 & 0.105 & 0.21 \\
\hline
\end{tabular}
\captionsetup{justification=centering}
\caption{Average $p$-values and rejection rates for the KS and CvM tests, using the resampling procedure for Scenario (HP).}  \label{table:comparison_ks_cvm}
\end{table}